\documentclass[longauth]{aa} 
\usepackage[labelfont=bf]{caption} 
\usepackage{graphicx}
\usepackage{subcaption} 
\usepackage{multirow}
\usepackage{booktabs}
\usepackage{natbib}
\usepackage{xcolor}
\usepackage{txfonts}
\usepackage{siunitx}
\usepackage{ulem}
\usepackage{threeparttable}

\begin{document}

   \title{Comparative analysis of missing data imputation methods for CSST survey: Impact on photometric redshift estimation performance
   }
   
\titlerunning{Imputation of missing photometric data and the evaluation of photo-$z$ accuracy for CSST}

   \author{Ling Wang\inst{1}
          \and
          Zhu Chen\inst{1}\fnmsep\thanks{Corresponding author: zhuchen@shnu.edu.cn}
          \and
          Zhijian Luo\inst{1}
          \and
          Liping Fu\inst{1,2}
          \and 
          Zuhui Fan\inst{3,4}
          \and
          Wei Du\inst{1}
          \and 
          Yaoming Lei\inst{1}
          \and
          Zhang Ban\inst{5}
          \and
          Yuedong Fang\inst{6}
          \and
          Yi Hu\inst{7}
          \and
          Xin Ji\inst{7}
          \and
          Guoliang Li\inst{8}
          \and
          Xiaobo Li\inst{5}
          \and
          Jiaqi Lin\inst{9,10,11}
          \and
          Chenxiaoji Ling\inst{7}
          \and
          Chao Liu\inst{7}
          \and
          Dezi Liu\inst{3}
          \and
          Changqing Luo\inst{7}
          \and
          Yu Luo\inst{12}
          \and
          Bin Ma\inst{9,11}
          \and
          Xianmin Meng\inst{7}
          \and
          Jundan Nie\inst{7}
          \and
          Juanjuan Ren\inst{7}
          \and
          Li Shao\inst{7}
          \and
          Jianing Tang\inst{13}
          \and
          Hao Tian\inst{7}
          \and
          Feng Wang\inst{14}
          \and
          Chengliang Wei\inst{8}
          \and
          Peng Wei\inst{7}
          \and
          Shoulin Wei\inst{15}
          \and
          Kaichao Wu\inst{13}
          \and
          You Wu\inst{7}
          \and
          Yun-Ao Xiao\inst{7,16}
          \and
          Zhou Xie\inst{14}
          \and
          Yibo Yan\inst{7,16}
          \and
          Su Yao\inst{7}
          \and
          Yan Yu\inst{7,9}
          \and
          Bo Zhang\inst{7}
          \and
          Shengwen Zhang\inst{7,16}
          \and
          Tianmeng Zhang\inst{7,16}
          \and
          Xiaoli Zhang\inst{13}
          \and
          Xin Zhang\inst{7,17}
          \and
          Bowei Zhao\inst{7,16}
          \and
          Zhimin Zhou\inst{7}
          \and
          Hu Zou\inst{7}
          }
\authorrunning{Wang}

   \institute{Shanghai Key Lab for Astrophysics, Shanghai Normal University, Shanghai 200234, China\and
              Center for Astronomy and Space Sciences, China Three Gorges University, Yichang 443000, People’s Republic of China\and
              South-Western Institute for Astronomy Research, Yunnan University, Kunming 650500, China\and
              Department of Astronomy, School of Physics,Peking University, Beĳing 100871, China\and
              Changchun Institute of Optics, Fine Mechanics and Physics, Chinese Academy of Sciences, Changchun, 130033, China\and
              University Observatory, Faculty of Physics, Ludwig-Maximilians-Universit\"at, Scheinerstr. 1, 81679 Munich, Germany\and
              National Astronomical Observatories, Chinese Academy of Sciences, 20A Datun Road, Chaoyang District, Beijing 100101, People's Republic of China\and
              Purple Mountain Observatory, Chinese Academy of Sciences, Nanjing, 210023, China\and
              School of Physics and Astronomy, Sun Yat-sen University, Zhuhai 519082, People's Republic of China\and
              Shanghai Astronomical Observatory, Chinese Academy of Sciences, 80 Nandan Road, Shanghai 200030, People's Republic of China\and
              CSST Science Center for the Guangdong-Hong Kong-Macau Greater Bay Area, Zhuhai 519082, People's Republic of China\and
              School of Physics and Electronics, Hunan Normal University, 36 Lushan Road, Changsha 410081, China\and
              Computer Network Information Center, Chinese Academy of Sciences, 2 East Kexueyuan South Road, Haidian District, Beijing 100083, People's Republic of China\and
              Center for Astrophysics and Great Bay Center of National Astronomical Data Center, Guangzhou University, Guangzhou, Guangdong 510006, People's Republic of China\and
              Faculty of Information Engineering and Automation, Kunming University of Science and Technology, No.727 Jingming South Road, Kunming, 650500, People's Republic of China\and
              School of Astronomy and Space Science, University of Chinese Academy of Sciences, Beijing 101408, People's Republic of China\and
              Key Laboratory of Space Astronomy and Technology, National Astronomical Observatories, Chinese Academy of Sciences, Beijing, 100101, China\\
             }

  \abstract{
    Improving the accuracy of photometric redshifts (photo-$z$) is essential for reliable statistical studies of cosmology and galaxy evolution. However, missing photometric bands are a common observational challenge that can significantly degrade photo-$z$ estimation accuracy. In this work, we present a systematic evaluation of data imputation methods aimed at improving photo-$z$ performance. We benchmark a range of representative machine learning (ML) and deep learning (DL) architectures, identifying k-nearest neighbors (KNN) and the attention-based SAITS model as the leading performers. These models are then applied to China Space Station Survey Telescope (CSST) mock data to assess their performance under realistic observational conditions. Our results show that KNN yields the highest accuracy under idealized missing completely at random (MCAR) conditions with complete training sets, whereas robustness tests reveal that SAITS significantly outperforms KNN when training data is incomplete or when applied to realistic mixed-mechanism scenarios. We find that domain consistency between training and testing missingness patterns is a prerequisite for optimal performance, highlighting the risks of domain shift in supervised regression tasks. Furthermore, our analysis demonstrates that while general imputation models are highly effective for MCAR and missing at random (MAR) data, they are detrimental when applied to missing not at random (MNAR) data arising from flux limits, as statistical models fail to capture the physical information inherent in these non-detections. Consequently, we advocate for more sophisticated architectures capable of disentangling stochastic missingness from physical non-detections to address these distinct mechanisms individually.
    }


 \maketitle

\section{Introduction}

Handling missing data is a common challenge in almost all fields of data analysis and has been extensively studied for decades \citep{Missing_Data}. Astronomical research is no exception, especially with regard to multi-band photometric data. Missing values originate from diverse sources, ranging from poor observing conditions and limitation of instruments to the non-detection of faint sources and artifacts introduced when cross-matching surveys of different depths and coverage. Consequently, the prevalence of such incomplete data presents a significant challenge to reliable model fitting and hinders advanced scientific inference. 

Three primary strategies are commonly used to address missing photometric data. The first and simplest is the case-wise deletion, which excludes any source with missing values from the sample. However, this approach has several drawbacks. First, it directly conflicts with the goal of maximizing the scientific utility of large astronomical surveys, as it inevitably reduces the sample size. Second, it risks introducing significant sample selection bias if the pattern of data missing is not completely random. 

The second strategy is to retain and flag missing values. This allows the algorithm with an inherent capability of handling missing data to proceed without explicit data imputation. For example, SED fitting algorithms \citep{Brammer_2008, 2006LePHARE, 2000HYPERZ} for photometric redshift (photo-$z$) can be configured to omit missing data points during $\chi^2$ minimization, while tree-based methods, such as random forest (RF) and boosting algorithms, can manage missing values intrinsically. However, the lack of complete information inevitably degrades the accuracy of downstream inferences like photo-$z$ regression. Furthermore, this strategy severely limits method selection, as many standard machine learning (ML) and deep learning (DL) models require complete numerical inputs. Similarly, traditional classification methods relying on color indices, used for star-galaxy separation or quasar selection \citep{2017Schindler}, fail if even a single necessary photometric band is unavailable.

The third and most sophisticated strategy is data imputation \citep{2013imputation}. This technique estimates missing values from observed correlations, typically using statistical models or ML algorithms. By reconstructing a complete dataset, successful imputation can significantly increase the effective sample size available for scientific analysis. Consequently, developing a robust imputation algorithm capable of generating reliable estimates represents the most promising approach for maximizing dataset utility and minimizing sample selection bias.

Traditional statistical methods for data imputation often rely on models with relatively simple distribution assumptions, and need to be re-engineered and validated for different data types, thereby limiting their universal application. In contrast, ML and DL methods are data-driven, offering a superior capacity to capture complex, high-dimensional data structures without predefined distributional forms, and making them far more versatile and widely applicable across diverse datasets. As AI techniques show powerful data mining ability on big data in recent years, several studies have begun to explore ML and DL for astronomical data imputation. 

An effective imputation algorithm must not only predict missing values with high precision but also preserve the underlying relationships between features (photometric data, colors, et al.) and the target labels (redshifts, object classification, et al.) after data completion. In the context of classification, \cite{missing2022} applied two variant imputation methods based on KNNimpute (k-nearest neighbor imputation; \citealt{ma2020_KNNimpute}) and LLSimpute (local least squares imputation; \citealt{wang_LLSimpute}) separately for the classification of transient events. The Euclid PHZ pipeline adopted the KNN algorithm \citep{knn} for data imputation before star-galaxy classification \citep{2025Euclid_PHZ}. Regarding photo-$z$ improvement, \cite{luken2021missing} compared conventional ML methods (KNN and MICE; \citealt{vanbuuren2000mice}) with a DL-based generative adversarial network (GAIN; \citealt{Yoon2018Gain}) using a relatively small ATLAS dataset of approximately 1300 objects, and found that MICE achieved the best performance. However, DL architectures are typically more complex than ML methods and generally require substantially larger training sets to avoid overfitting and to learn robust feature representations. Consequently, a sample of this size may limit the effectiveness of the DL model and may not provide a fully representative benchmark for comparing DL- and ML-based approaches. \cite{luo_paper} implemented GAIN on a large CSST simulation dataset, and demonstrated its superior performance. \cite{2023ApJ_Chartab} used the RF model to predict the magnitude of drop bands in their datasets, however, as their main goal was to apply information theory to reduce redundancy in band selection for optimal parameter prediction, they did not perform an extensive validation of the imputation itself. \cite{2024AJ_La} constructed an SOM likelihood from complete training data to infer the color of galaxies with missing data, thereby maximizing the data utility to improve the statistical accuracy of galaxy parameters. This strategy integrates DL with probabilistic techniques, relying on the assumption that the color distributions remain consistent between the training and test datasets.

Despite the development of various ML and DL methods for photometric data imputation, there is currently no consensus or clear guidance on which of these frameworks performs optimally for the unique characteristics of astronomical multi-photometry data. Given the significant advancements and robust performance of existing general-purpose ML/DL techniques, understanding their utility and limitations within our specific domain is warranted, rather than focusing on the creation of entirely new imputation frameworks. Consequently, a comprehensive and systematic evaluation of representative models spanning diverse architectural families is essential to elucidate the most effective strategies for addressing the challenges inherent to astronomical multi-photometry imputation. In this study, we systematically evaluate imputation performance by assessing the accuracy of photo-$z$ estimations derived from a range of representative models.

Precise galaxy redshifts are fundamental for deriving key physical properties, such as mass and luminosity, and for studying the large-scale structure and evolution of the universe \citep{evolution1, evolution, Mo, Abdalla2011}.  The gold standard for accuracy is the spectroscopic redshift, which is determined by the analysis of discrete spectral features \citep{specz1, specz2, specz3}. However, this spectroscopic method is observationally expensive and often requires long exposure times to achieve an adequate signal-to-noise ratio (S/N) \citep{spec_SN}, making it impractical for very large samples. This limitation has necessitated the widespread adoption of photo-$z$ as a less precise but far more efficient alternative \citep{photoz1, photoz2, Feldmann}. By estimating redshifts from broad- and medium-band photometry, this technique is pivotal for modern multi-band sky surveys, facilitating cosmological statistical studies on galaxy populations that are orders of magnitude larger than those accessible to spectroscopy. As the era of large-scale surveys ushers in unprecedented opportunities for precision cosmology, achieving scientific goals remains critically dependent on deriving high-precision photo-$z$. Such accuracy is a cornerstone for major ongoing missions, including the Dark Energy Camera Legacy Survey (DECaLS; \citealt{2019DECaLS}), the Vera C. Rubin Observatory’s Legacy Survey of Space and Time (LSST; \citealt{2009LSST, 2019LSST}), the Euclid mission \citep{2011Euclid}, and also essential for the forthcoming survey of Chinese Space Station Survey Telescope (CSST) \citep{1CSST, 2CSST, CSST_2025}. In this paper, we use simulated data for the CSST survey to systematically evaluate and compare the performance of several imputation techniques in improving photo-$z$ precision.

The remainder of this paper is organized as follows. Section 2 details the simulation dataset and the methodology used to generate realistic missing data patterns. Section 3 introduces the suite of imputation models under consideration. Our core comparative analysis is presented in Section 4, where we systematically evaluate the models based on the resulting photo-z accuracy from EAZY \citep{Brammer_2008} and assess their stability to select optimal approaches. In Section 5, we apply the optimal methods to realistic missing data scenarios anticipated for CSST. Finally, Section 6 summarizes our key findings and presents the conclusions of this study.

\begin{figure*}[!t]
        \centering
    \includegraphics[width=1\textwidth]{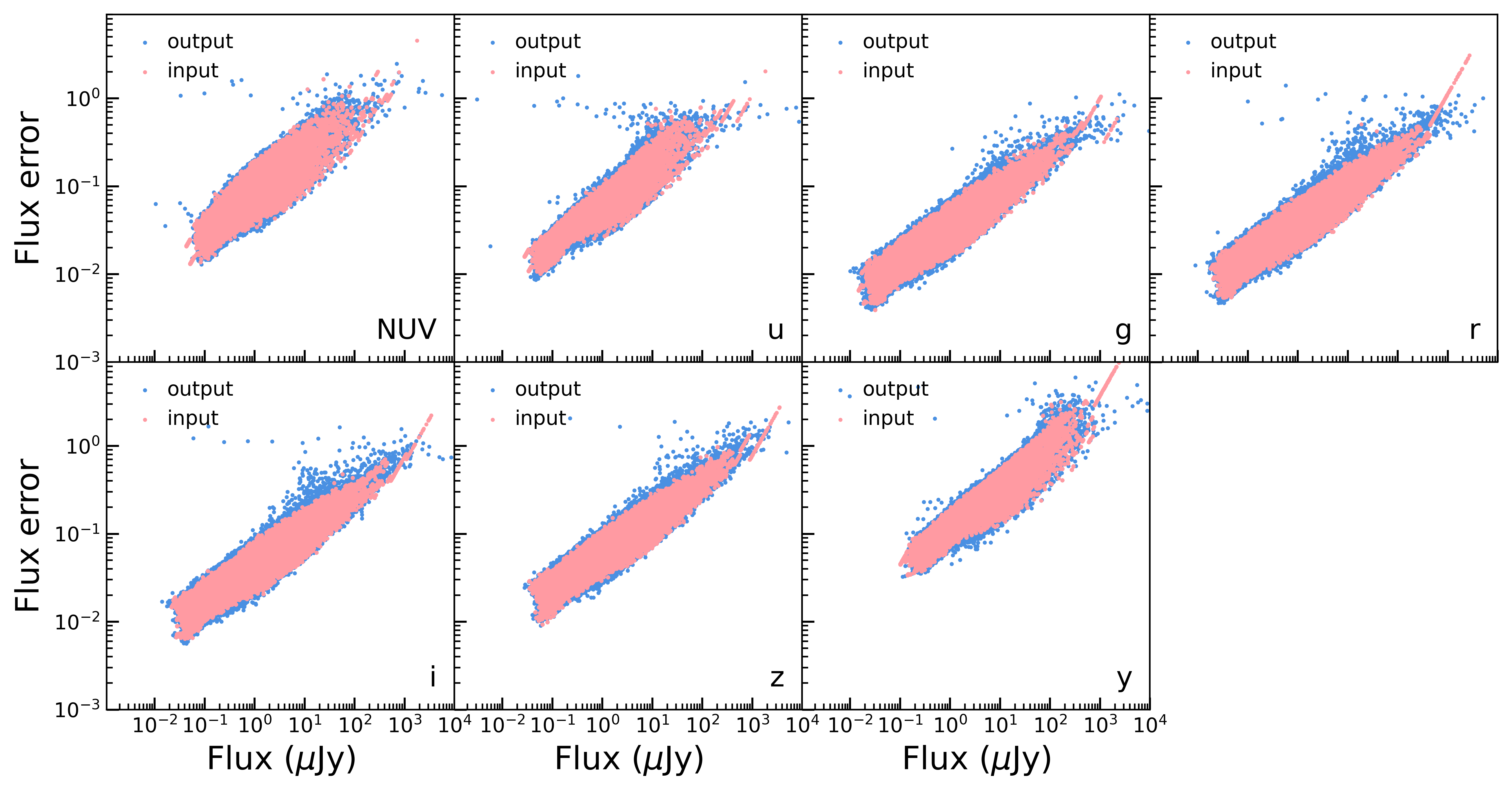}
    \caption{Flux–error relations across bands. Input flux vs. predicted flux errors (pink) and output catalog values (blue). 
}
    \label{fig:fluxerro}
    \end{figure*}

\section{Dataset and evaluation method}

CSST is a 2-meter space telescope designed to be co-orbital with the China Manned Space Station \citep{CSST_2025}. The CSST Survey Camera plans to conduct cumulative observations for approximately 7 years during the 10-year orbital period, obtaining wide-field survey data of 7-band ($NUV$, $u$, $g$, $r$, $i$, $z$, and $y$) photometric imaging over a sky area of about 17 500 deg$^2$. The survey will reach 5$\sigma$ point-source depths of approximately 26 AB mag in the $g$, $r$, and $i$ bands, and 24.4–25.4 AB mag in the remaining bands. The ambitious scientific goals of CSST include probing the evolution of large-scale structures, constraining the nature of dark matter and dark energy, and investigating galaxy formation and evolution \citep{csst, csst1, csst2, 2023_liu}. As China’s first optical space telescope, the CSST pipeline team has developed comprehensive simulations to model the survey strategy and instrumental effects \citep{Wei_2026_024001, Ban_2026, Wei_2026_024004, Xian_2026}. These efforts are crucial for building the data processing pipeline and evaluating the scientific potentials of the current survey design.

Our evaluation is based on data from the CSST Cycle 6 survey simulation, which emulates a 1.53 deg² observation centered at R.A. = 244.97° and Dec = 39.90°. The simulation is designed for high fidelity, incorporating both physical phenomena like weak lensing and cosmic rays, as well as instrumental effects such as the point spread function, dark current, bias, flat-fielding, and various detector artifacts. The underlying galaxy population for this simulation is sourced from the comprehensive Jiutian cosmological simulation \citep{jiutian_2025}, which provides ground-truth properties for objects down to a magnitude of 28 across redshift up to 2.0.

This simulation yields two primary, co-spatial catalogs. The first is the input catalog as ground truth, which contains the intrinsic properties of all sources and serves as our benchmark for training and evaluation. The second is the output photometry catalog, a Level 2 data product generated by running the official CSST pipeline on the simulated images. This catalog contains realistic photometry and, critically, inherits authentic data missing patterns from the observational and processing limitations. It therefore represents the incomplete dataset that our imputation methods must address.

\subsection{Dataset construction}

To establish a reliable ground truth for our analysis, we first constructed a complete dataset by selecting sources from the simulated CSST input catalog that possess valid photometric measurements across all seven bands.

However, the input catalog only provides true, error-free magnitudes, whereas photometric uncertainties are unavoidable in actual observations. To produce a more realistic dataset suitable for downstream tasks such as SED fitting, which require flux errors, we implemented a magnitude error prediction model. This model leverages a RF algorithm, specifically trained to match the noise properties of the seven CSST bands. It operates by predicting realistic flux errors through a S/N matching mechanism, subsequently applying these errors to the true magnitudes in our complete input catalog.

The success of this error modeling is illustrated in Figure \ref{fig:fluxerro}. We show the flux-error correlation for both the actual simulated pipeline output and our error-modeled input catalog. The close alignment between these two distributions shows that our model effectively reproduces the observational characteristics of the S/N, thus improving the realism of our complete dataset.

\begin{table}[!t]
\centering
\caption{Dataset partitioning and missing data settings.}
\label{tab:Datasets.}
    \resizebox{\linewidth}{!}{
	\begin{tabular}{llll}
		\toprule
		Dataset & Sample size & Drop bands & Drop rate(\%) \\
		\midrule
		Test set & 30 000 & 1, 2, 3 & 10, 20, 30 \\
		Validation set & 30 000 & 1, 2, 3 & 10, 20, 30 \\
		Training set & 5000, 10 000, 20 000, 30 000,  & full & 10, 20, 30 \\
        & 60 000, 90 000, 120 000 & & \\
		\bottomrule
	\end{tabular}
    }
\end{table}

To systematically evaluate our imputation methods, we divide the complete dataset into distinct training, validation, and test sets. These divisions serve standard roles: the training set for model fitting, the validation set for hyperparameter tuning and preventing overfitting, and the test set for an unbiased final performance assessment. The details of these datasets are provided in Table \ref{tab:Datasets.}.

Our experimental design for the training sets was twofold. First, to rigorously assess the impact of sample size on imputation performance, we constructed seven nested training sets of varying sizes. This nested approach ensures that each larger set is a superset of the smaller ones, thus isolating the effect of sample size from potential variations due to random sampling. Second, to investigate the influence of missingness in the training data itself, we introduced various missing data fractions into a specific 120 000 sample training set.

The test sets were specifically designed for a comprehensive and unbiased evaluation of the trained models under various conditions, and their corresponding validation sets were constructed with identical data missing patterns to ensure consistent evaluation during model development. We employed two schemes to introduce missing values under the missing completely at random (MCAR) assumption:

\begin{itemize}
\item Scenario-based missingness: to assess model performance across a spectrum of common missingness patterns, we created test scenarios by selectively removing different combinations of photometric bands (e.g., one, two, or three bands per source).

\item Robustness-based missingness: to rigorously evaluate model robustness against varying degrees of data incompleteness, we applied a range of global missing rates (10\%, 20\%, or 30\% of all photometric measurements) across the test set. The overall missing rate is defined as the total number of missing entries divided by the total number of possible entries in the dataset.
\end{itemize}

Following the imputation of missing magnitudes, a critical subsequent step is the generation of their corresponding flux errors. Since a missing magnitude inherently implies a missing error, we re-apply our previously described RF error model to the newly imputed magnitudes. This ensures that every entry in the completed catalog possesses both a magnitude and a realistic error, a prerequisite for photo-$z$ estimation.

Our decision to employ the MCAR mechanism in this initial evaluation warrants explanation. While real observational data, such as CSST, will undoubtedly exhibit a complex mix of missing at random (MAR) and missing not at random (MNAR) components (e.g., non-detections of faint objects), our objective here is to establish a baseline performance comparison for imputation models within a controlled, idealized environment \citep{review, VanBuuren_2018}. Therefore, this initial analysis focuses exclusively on the MCAR case, with a comprehensive investigation into MAR and MNAR imputation deferred to future work.

\subsection{Photo-$z$ evaluation metrics}

The ultimate application of an imputation method is its ability to improve downstream scientific results. Given that incomplete photometric data inherently degrade the reliability of photo-$z$ determinations, we evaluate photo-$z$ performance as our principal criterion for assessing imputation quality. We assess this performance using three key metrics.

The first metric is outlier fraction (\( f_{\text{out}} \)). This metric identifies the fraction of sources with catastrophic photo-$z$. An object is considered as an outlier if its estimated photo-$z$ deviates significantly from its true redshift (\( z_{\text{in}} \)), per the condition in
\begin{equation}
\frac{|z_{\text{in}} - z_{\text{phot}}|}{1 + z_{\text{in}}} > 0.15,
\end{equation} 
the outlier fraction is $f_{\text{out}} = \frac{N_{\text{outlier}}}{N_{\text{total}}}$ \citep{metric_outlier}.

The second metric is the normalized median absolute deviation (\(\sigma_{\text{NMAD}}\)), which provides a robust measure of the photo-$z$ precision (i.e., scatter). Following \citet{Brammer_2008}, it is defined as
 \begin{equation}
\sigma_{\text{NMAD}} = 1.48 \times \text{median} \left( \frac{\Delta z - \text{median}(\Delta z)}{1 + z_{\text{in}}} \right),
\end{equation}
where $\Delta z = z_{\text{phot}} - z_{\text{in}}$. The scaling factor of 1.48 makes this statistic comparable to the standard deviation of a normal distribution.

The final metric is the bias of the photo-$z$ (\textit{bias}), revealing whether redshifts are generally overestimated or underestimated. It is calculated as the median of the normalized residuals
\begin{equation}
\textit{bias} = \mathrm{median} \left( \frac{z_{\mathrm{in}} - z_{\mathrm{phot}}}{1 + z_{\mathrm{in}}} \right),
\label{eq:bias}
\end{equation}

All photo-$z$ were derived using the EAZY code \citep{Brammer_2008}. Since different template-fitting codes tend to produce statistically comparable photo-$z$ accuracy when run with the same configuration \citep{Desprez2020}, our findings using EAZY are expected to be broadly applicable. Our specific configuration utilized the "tweak\_fsps\_QSF\_v12\_v3" templates, which is based on the flexible stellar population synthesis (FSPS) code \citep{tweak_templets}, a standard $\Lambda$CDM cosmology ($H_0 = 70\ \mathrm{km}\ \mathrm{s}^{-1}\ \mathrm{Mpc}^{-1}$, $\Omega_{\mathrm{m}} = 0.3$, $\Omega_{\Lambda} = 0.7$), and default values for all other parameters.

\section{Imputation methods}

In selecting models for evaluation, we made a deliberate choice to prioritize methods representing fundamentally different underlying architectures. The overall accuracy and suitability of an imputation algorithm for a given dataset, particularly in complex domains like astronomical multi-photometry, are primarily driven by its basic architectural paradigm rather than by minor variations or specific implementations within a single family of models. While iterative refinements and hyperparameter tuning can yield incremental improvements, significant performance differences are more likely to arise from the core structural approach (e.g., whether it’s distance-based, sequence-aware, or attention-based). Therefore, our selection criteria emphasized models built upon diverse architectural foundations to identify the most promising foundational approaches for this application.

\subsection{Machine learning methods}

\paragraph{KNN (k-nearest neighbors)} 
The KNN algorithm \citep{knn} is a distance-based learning method. For imputation, it identifies the k-nearest neighbors for a sample with missing values based on a distance metric (e.g., Euclidean distance) computed from the available observed features. The missing values are then estimated by a weighted average or the mean of the corresponding feature values from these neighbors. In this study, we implement KNN imputation using the KNNImputer module from the Scikit-learn library \citep{SKlearn_2011}.

\paragraph{RF}
RF \citep{2001RF} is an ensemble learning algorithm that performs regression by constructing a multitude of decision trees. For imputation, each tree is trained on a bootstrapped subsample of the data. To predict a missing value for a specific band, the RF model is trained to predict that band’s magnitude using the other observed bands as features. The final imputed value is the average of the predictions from all trees in the forest, which enhances robustness and reduces overfitting. The optimal number of trees is determined via k-fold cross-validation. The RF model we adopted is also implemented based on the Python interface of the Scikit-learn library \citep{SKlearn_2011}.

\paragraph{CatBoost (categorical boosting)}
CatBoost \citep{2018CatBoost, 2018_Veronika} is a high-performance gradient boosting algorithm that builds decision trees sequentially, with each new tree correcting the errors of its predecessor. While primarily known for its sophisticated handling of categorical features, CatBoost also offers a powerful built-in mechanism for handling missing numerical values (via the nan\_mode hyperparameter). This allows it to directly process and impute missing data without requiring external pre-processing steps, making it a highly efficient and integrated solution. The imputation is performed as part of the model’s internal training process when predicting a target variable. In terms of CatBoost, we directly employed its official Python package for our experiments \footnote{https://github.com/catboost}.

\subsection{Deep learning methods}

In recent years, DL has emerged as a powerful approach for missing value imputation. In this study, we utilize the PyPOTS Python package \citep{du2023pypots} \footnote{https://github.com/WenjieDu/PyPOTS}, a comprehensive library for partially-observed time series data mining, especially for data imputation. We select a suite of state-of-the-art DL models designed for time-series data, also suitable for sequential data from the package, which are categorized by their underlying architectures below.

\paragraph{RNN-based}
These models utilize recurrent neural networks to model sequential dependencies. We selected M-RNN (missing data recurrent neural network; \citealt{MRNN})  and BRITS (bidirectional recurrent imputation for time series; \citealt{BRITS}) for evaluation. M-RNN contains both an interpolation block and a subsequent imputation block, which are trained jointly. The model uses the interpolated values as an initial guess to improve the accuracy of the final imputation within the RNN structure. BRITS improves upon earlier RNN-based methods by treating imputed values as learnable variables within a bidirectional recurrent system. This allows the imputed values to be directly updated via backpropagation, leading to more accurate estimates compared to models like M-RNN where imputed values are treated as fixed constants during training updates. A notable drawback of many RNN-based models, however, is their autoregressive nature, which can lead to error accumulation \citep{Venkatraman2015}.

\paragraph{GAN-based}
These models frame imputation as a generative task, learning the underlying data distribution. We adopted US-GAN (unsupervised generative adversarial network; \citealt{usgan}) as our representative model. US-GAN adapts the standard GAN architecture for imputation, consisting of a Generator that learns to impute missing values and a Discriminator that attempts to distinguish between real and imputed data. The Generator, typically a Bidirectional RNN, is trained to minimize both an adversarial loss (to fool the Discriminator) and a reconstruction loss (to ensure consistency with observed data). 

\begin{figure*}[!t]
    \centering
    \begin{subfigure}{1\linewidth}
	\centering
    \includegraphics[width=1\textwidth]{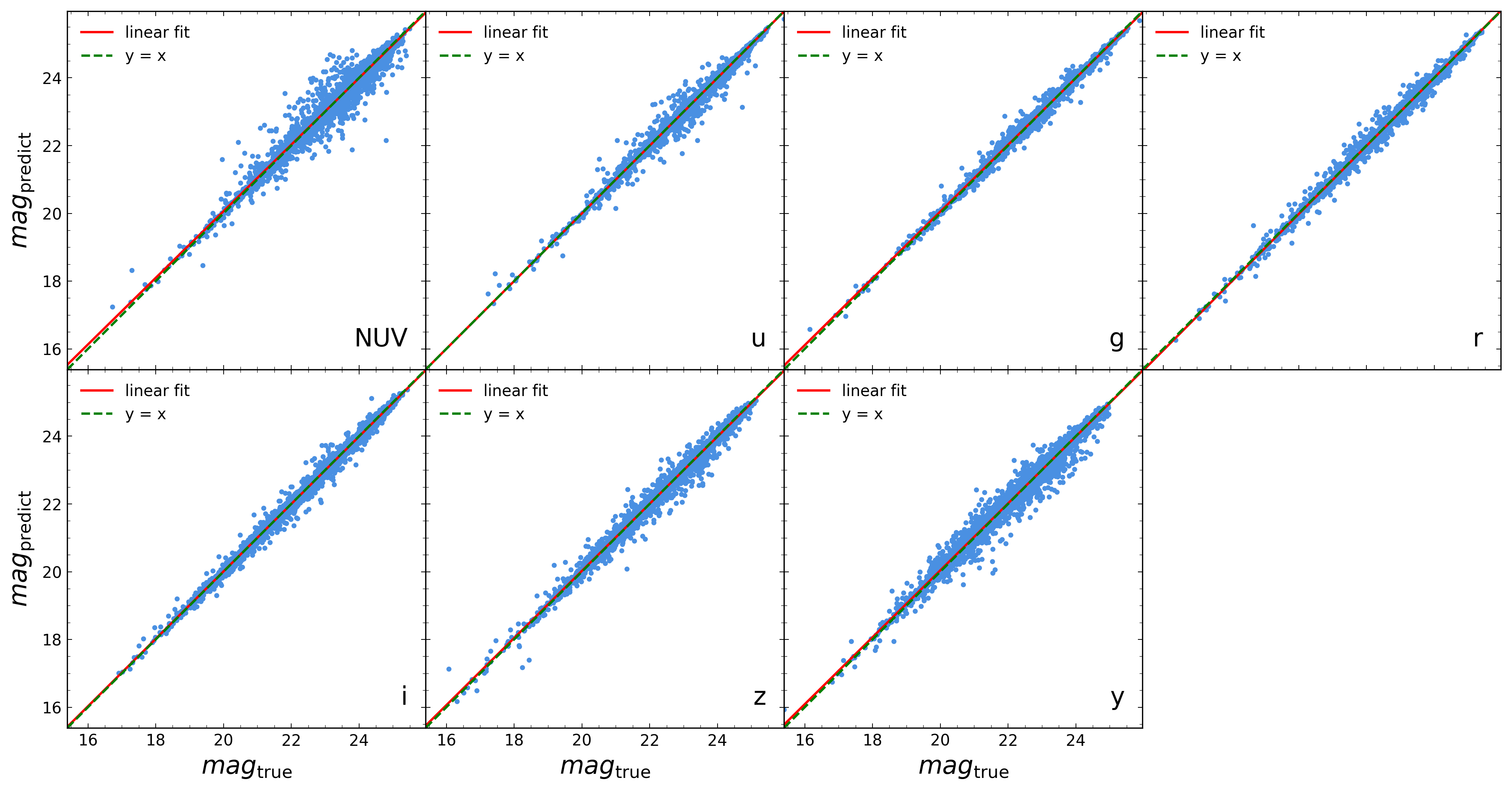}
    \end{subfigure}
    \begin{subfigure}{1\linewidth}
	\centering
    \includegraphics[width=1\textwidth]{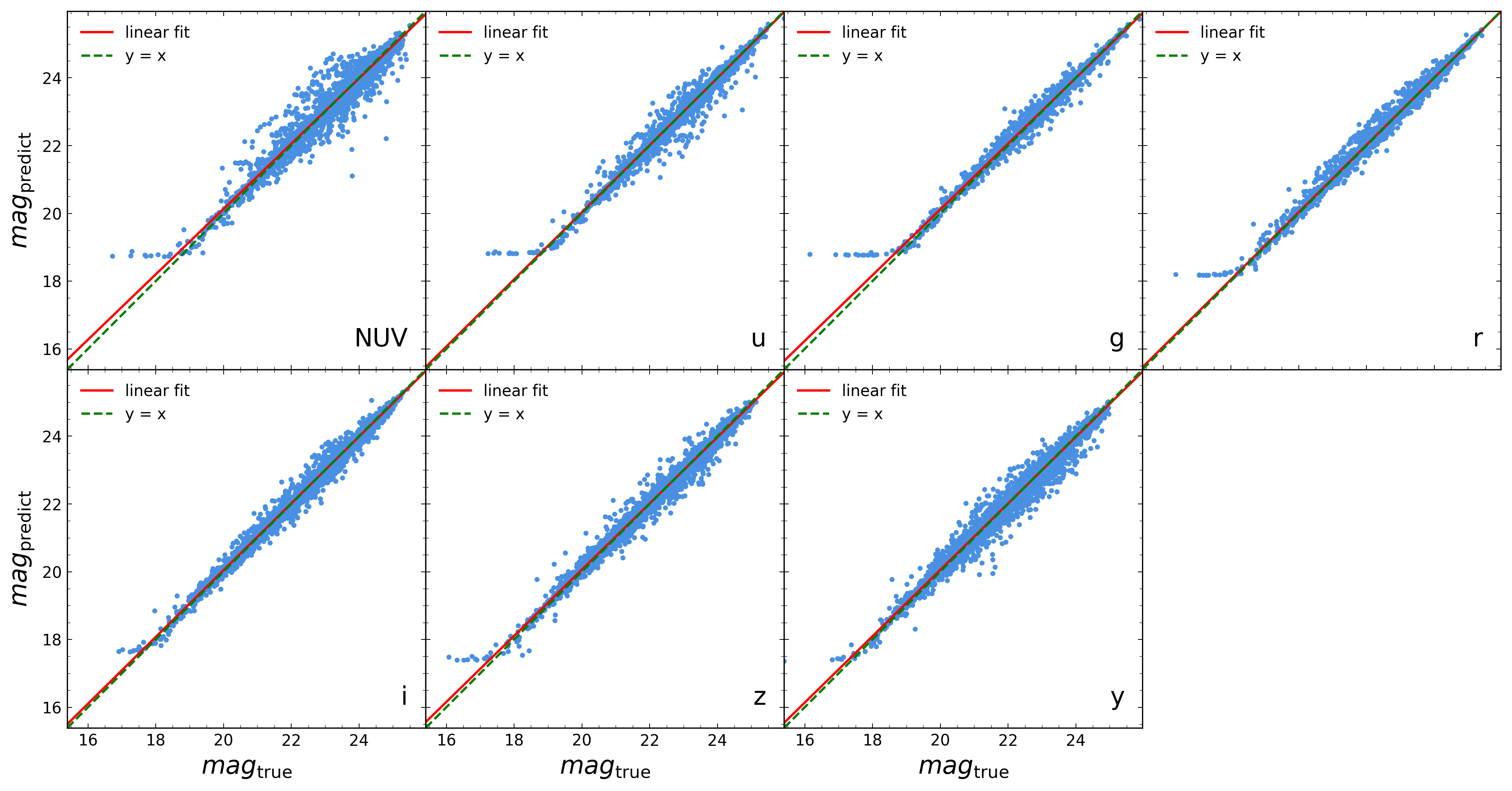}
    \end{subfigure}
    \caption{Predicted versus true magnitudes for the test set with three missing photometric bands using the imputation models. Panel a: Predicted versus true magnitudes (KNN). Panel b: Same as (a) but for the SAITS model. We show the predicted values as a function of the true values. The dashed green line marks $\mathit{mag}{\mathrm{predict}}=\mathit{mag}{\mathrm{true}}$, and the red line shows the linear fit.}
    \label{fig:input_imputation_dropband}
\end{figure*}
 
\paragraph{VAE-based}
These models use a variational autoencoder framework to learn a low-dimensional latent representation of the data. We chose GP-VAE (gaussian process variational autoencoder; \citealt{GP-VAE}) , a model specifically designed for time series with missing values. GP-VAE maps the incomplete high-dimensional data to a continuous, low-dimensional latent space where there are no missing values. Crucially, it places a Gaussian Process prior on this latent space, enforcing temporal smoothness and structure. The imputation is performed by encoding the incomplete data into this latent space and then decoding it back to the original data space, effectively reconstructing the missing values based on the learned temporal dynamics.

\paragraph{Self-Attention-based}
This category of models replaces traditional recurrence with attention mechanisms to capture dependencies across the entire data sequence. We included the original Transformer model \citep{Transformer}, a recent variant iTransformer \citep{itransformer}, and SAITS (self-attention imputation for time series; \citealt{SAITS}) as test models. The original Transformer model relies entirely on multi-head self-attention to capture global dependencies between input and output. While designed for sequence transduction, its ability to relate different positions of a single sequence makes it applicable to imputation tasks. iTransformer inverts the standard Transformer architecture. It applies self-attention across different variables (i.e., photometric bands) at the same time step, treating the variables themselves as tokens. This is particularly well-suited for capturing the multivariate correlations inherent in photometric data. The SAITS model is specifically designed for imputation. It employs a joint-optimization objective that combines a standard imputation loss with a reconstruction loss. Its core consists of two diagonally-masked self-attention blocks, which explicitly capture both temporal dependencies and feature correlations to effectively impute missing values in multivariate time series.

\section{Experiments and results}\label{Model imputation capability experiments}
\begin{table*}[!t]
  \begin{threeparttable}
  \centering
  \caption{Performance comparison of different methods on various drop bands. 
  }
\small 
  \renewcommand{\arraystretch}{1.2}
  \setlength{\tabcolsep}{2.6pt} 
			\begin{tabular}{ccccccc}
				\hline
				\multirow{2}{*}{\textbf{Model}} & \multicolumn{2}{c}{\textbf{Drop 1 band}} & \multicolumn{2}{c}{\textbf{Drop 2 bands}} & \multicolumn{2}{c}{\textbf{Drop 3 bands}} \\ 
                \cmidrule(r){2-3}  \cmidrule(r){4-5} \cmidrule(r){6-7}
				& $f_{out}$(\%)/$\sigma_{\text{NMAD}}$/bais & mae/rmse/mre & $f_{out}$(\%)/$\sigma_{\text{NMAD}}$/bais & mae/rmse/mre & $f_{out}$(\%)/$\sigma_{\text{NMAD}}$/bais & mae/rmse/mre \\ \hline
                Test (non-imputed) & 3.79/0.061/0.0022 & ... & 11.33/0.086/-0.0112 & ... & 26.34/0.131/-0.0375 & ... \\ 
                \hline
				KNN & 1.41/0.049/0.0100 & 0.028/0.062/0.036 & 1.73/0.051/0.0099 & 0.033/0.073/0.041 & 2.48/0.053/0.0095 & 0.040/0.088/0.050 \\
				RF & 1.45/0.049/0.0100 & 0.027/0.060/0.034 & 7.38/0.059/0.0050 & 0.124/0.291/0.158 & 19.22/0.080/0.0038 & 0.236/0.443/0.299 \\
				CatBoost & 1.73/0.051/0.0094 & 0.048/0.099/0.061 & 18.17/0.077/0.0148 & 0.258/0.383/0.329 & 34.63/0.129/0.0289 & 0.364/0.527/0.463 \\ 
				US-GAN & 4.09/0.055/0.0087 & 0.182/0.252/0.231 & 18.88/0.080/0.0308 & 0.325/0.420/0.410 & 21.02/0.094/0.0062 & 0.277/0.397/0.351 \\ 
				GP-VAE & 7.97/0.064/0.0152 & 0.231/0.304/0.293 & 22.78/0.093/0.0243 & 0.309/0.393/0.391 & 32.26/0.123/0.0032 & 0.390/0.517/0.494 \\
				M-RNN & 32.79/0.107/0.0285 & 0.785/1.013/0.995 & 44.93/0.175/0.0411 & 0.787/1.004/0.995 & 51.22/0.215/0.0543 & 0.785/1.005/0.994 \\ 
				BRITS & 4.04/0.054/0.0087 & 0.163/0.289/0.207 & 8.94/0.065/0.0147 & 0.210/0.344/0.256 & 20.02/0.094/0.0229 & 0.257/0.373/0.326 \\ 
				Transformer & 1.94/0.052/0.0104 & 0.071/0.119/0.090 & 2.58/0.053/0.0093 & 0.058/0.114/0.074 & 5.72/0.063/0.0069 & 0.091/0.151/0.115 \\ 
				iTransformer & 1.67/0.052/0.0095 & 0.051/0.103/0.064 & 2.40/0.054/0.0100 & 0.053/0.109/0.067 & 4.37/0.060/0.0074 & 0.075/0.139/0.095 \\ 
				SAITS & 1.41/0.048/0.0100 & 0.024/0.055/0.030 & 2.10/0.052/0.0106 & 0.043/0.087/0.054 & 3.27/0.056/0.0080 & 0.051/0.107/0.065 \\ 
				\hline
			\end{tabular}
		\label{tab:performance}
\tablefoot{The $f_{\mathrm{out}}$, $\sigma_{\mathrm{NMAD}}$, and bias values 
for the complete test sample are 1.33\%, 0.048, and 0.0104, respectively. These quantities represent the standard photo-$z$ evaluation metrics. The lower, the better.}
    
  \end{threeparttable}
\end{table*}

\begin{table*}[!t]
	\centering
	\caption{Number of model parameters and training time per epoch for different numbers of missing bands.}
      \setlength{\tabcolsep}{17.5pt} 
	\renewcommand{\arraystretch}{1.2}
	\renewcommand{\tablename}{Table}
	
	\begin{tabular}{c cc cc cc}
		\hline
		\multirow{2}{*}{\textbf{Model}} 
		& \multicolumn{2}{c}{\textbf{Drop 1 band}} 
		& \multicolumn{2}{c}{\textbf{Drop 2 bands}} 
		& \multicolumn{2}{c}{\textbf{Drop 3 bands}} \\ 
		\cmidrule(r){2-3} \cmidrule(r){4-5} \cmidrule(r){6-7}
		& \# of param & s/epoch 
		& \# of param & s/epoch 
		& \# of param & s/epoch \\ 
		\hline
		US-GAN        & 140 745  & 17.83 & 79 295  & 17.06 & 106 607 & 17.03 \\ 
		GP-VAE        & 40 897   & 6.69  & 143 041 & 12.65 & 34 273  & 6.89 \\ 
		M-RNN         & 35 231   & 4.98  & 50 381  & 5.18  & 20 615  & 5.32 \\ 
		BRITS         & 46 636   & 15.21 & 75 826  & 15.64 & 75 826  & 15.35 \\
		Transformer   & 331 521  & 12.96 & 405 697 & 12.48 & 864 033 & 12.72 \\
		iTransformer  & 811 015  & 11.48 & 924 807 & 10.91 & 925 703 & 10.99 \\ 
		SAITS         & 1 713 677& 12.95 & 1 056 781& 12.23 & 1 873 677& 12.68 \\ 
		\hline
	\end{tabular}
    \tablefoot{Training time is given in seconds. The default maximum number of epochs is 100.
    }
	\label{tab:model_params}
\end{table*}

To evaluate model performance, we used a fixed-size training set of 120 000 complete sources. The test sets were generated by randomly removing one, two, or three bands from each source. We assessed imputation quality using three standard metrics: mean absolute error (MAE), root mean square error (RMSE), and mean relative error (MRE). For a fair comparison, all models were trained with a batch size of 1024. This relatively large batch size was chosen due to the low dimensionality of the input data and the moderate model size, which allow efficient utilization of GPU memory. In practice, we found that larger batch sizes led to more stable training behavior. Hyperparameters were tuned using Optuna (Akiba et al. 2019), a Bayesian optimization framework, where the learning rate was treated as a tunable parameter to account for its coupling with batch size. An early stopping strategy was employed, halting training if the validation MAE did not improve for 5 consecutive epochs, all models converged within 100 epochs. All experiments were conducted on a single NVIDIA RTX 3090 GPU.
    
\subsection{Imputation performance comparison}\label{imputation_performance}

Table \ref{tab:performance} reports the imputation performance of models on the three test datasets, as well as the accuracy of photo-$z$ estimation for the imputed photometric datasets. Across all test scenarios, two models consistently emerged as the top performers: KNN and, to a close second, SAITS. Both models dramatically improve photo-$z$ accuracy, approaching the performance achieved with the complete, original dataset. The impact is particularly striking in the most challenging case with three missing bands, where both methods reduce the photo-$z$ outlier fraction by a factor of 10 compared to the non-imputed baseline. The high fidelity of these imputations is visually confirmed in Figure \ref{fig:input_imputation_dropband}, which compared the true versus imputed magnitudes. Figure \ref{fig:input_imputation_dropband}a shows that KNN provides exceptionally accurate and unbiased imputations, while SAITS also performs very well (Figure \ref{fig:input_imputation_dropband}b), it exhibits a slight systematic bias at the bright end, which we attribute to the relative scarcity of bright sources in the training data. To demonstrate the practical utility of our top-performing models, we also applied them to the more realistic output catalog, which incorporates observational noise and pipeline processing artifacts. For this purpose, the models were retrained using a subsample of the output catalog with complete data, enabling adaptation to additional observational features that are not present in the input catalog. Both KNN and SAITS again demonstrated robust performance. This is confirmed in Figure \ref{fig:output_imputation_dropband} in Appendix \ref{appendix1}, which also shows the successful imputation of the output catalog even in the challenging scenario of three missing bands. 

It is also noteworthy that while all tested ML methods performed well with a single dropped band, performance diverged as data sparsity increased. Tree-based methods (RF and CatBoost) exhibited a rapid decline in photometric accuracy as the number of missing bands rose, whereas KNN—a distance-based method—proved to be more robust. A similar trend was observed in DL models: RNN-based and GAN-based methods failed to maintain robustness as the number of missing bands increased, while transformer-based methods demonstrated significantly more stable performance. 

\begin{figure*}[t]
    \centering
    \includegraphics[width=1\textwidth]{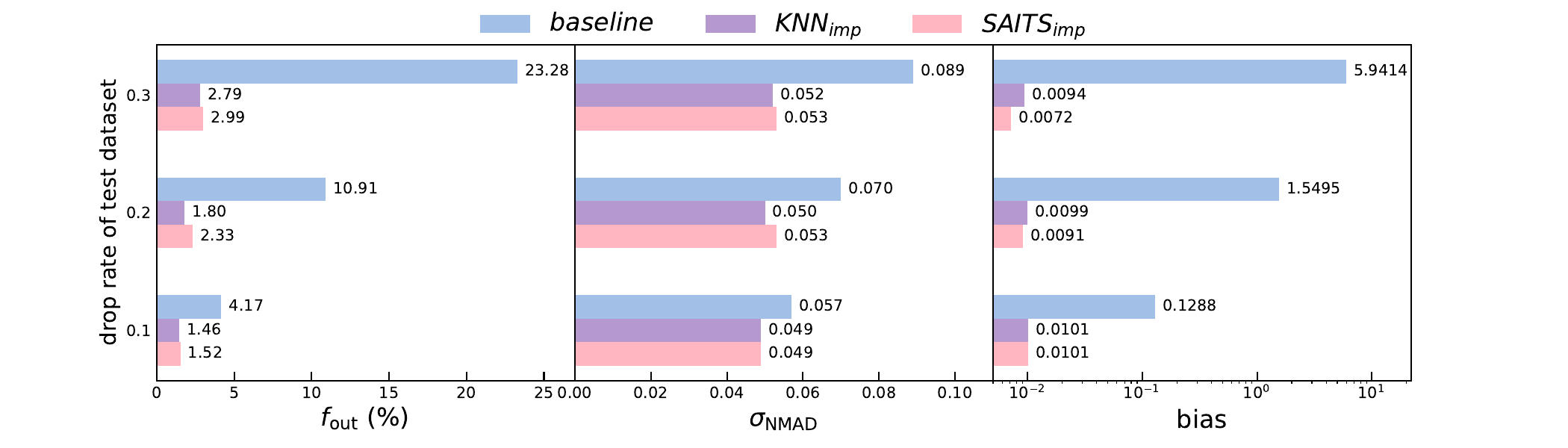}
    \caption{Photo-$z$ quality metrics for the KNN and SAITS models for test sets with different missing data rates. Blue bars show SED fitting on non-imputed data, and pink and purple bars show results after KNN and SAITS imputation, respectively.}
    \label{fig:Different_test_drop_rate}
\end{figure*}

\begin{figure}[t]
    \centering
    \includegraphics[width=0.5\textwidth]{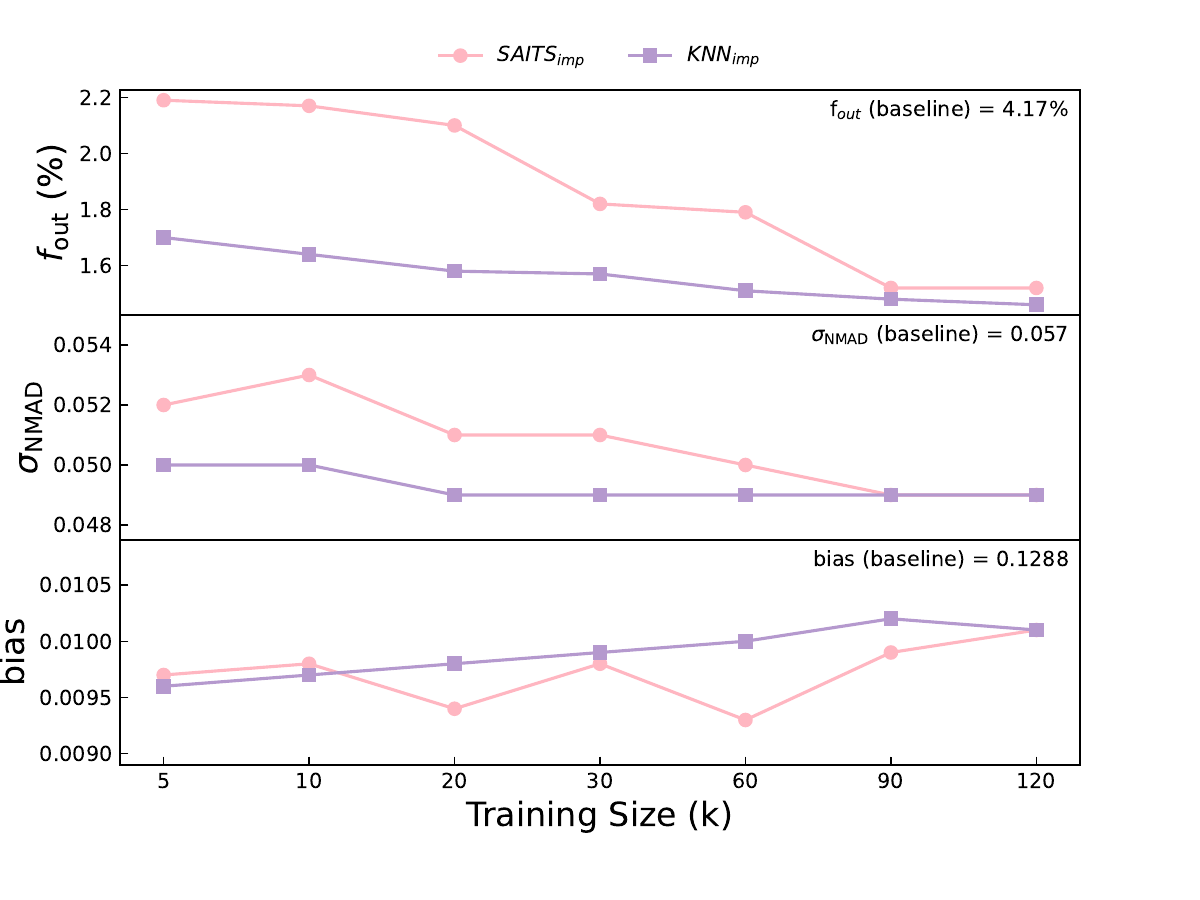}
    \caption{Influence of training sample size on imputation performance. We show the photo-$z$ metrics after imputation with the SAITS (pink) and KNN (purple) models for different training sample sizes. The non-imputed test set has a missing rate of 10\%, and the corresponding metrics are shown in the upper right corner of each panel.}
    \label{fig:different_training_size}
\end{figure}

Beyond imputation accuracy, computational cost is a critical practical consideration. Table \ref{tab:model_params} summarizes the parameter counts and training speeds for the DL models. Although transformer-based methods are more complex and contain more parameters than other DL architectures, their training time per epoch remains acceptable and is justified by the substantial improvement in photo-$z$ accuracy. It is worth noting that the reported parameter counts vary across different drop-band scenarios, as the hyperparameters of each model are independently optimized using Optuna for each setting. Consequently, the resulting model complexity does not necessarily exhibit a monotonic relationship with the number of missing bands. This can be attributed to the stochastic nature of hyperparameter optimization in a high-dimensional search space.

\subsection{Robustness analysis}	

Our previous experiments identified KNN and SAITS as the top-performing imputation models. However, realistic astronomical catalogs exhibit varying degrees of data sparsity, and relying on large, complete training sets is often impractical and prone to selection bias. Therefore, in this section, we conduct a rigorous evaluation of the robustness of these two leading models under variable conditions. Specifically, we assess their photo-$z$ accuracy by systematically varying three key scenarios: the sparsity of the test set, the sample size of the training set, and the degree of incompleteness within the training data.

\subsubsection{Drop rate of test dataset analysis}\label{Different drop rate}	

First, we evaluated the models’ robustness to varying degrees of missing data in the test set. For this experiment, we retained the training dataset described in Section \ref{imputation_performance} but generated three separate test datasets with global missing rates of 10\%, 20\% and 30\%, as detailed in Table \ref{tab:Datasets.}. For hyperparameter tuning, we utilized validation datasets with missing rates matching their corresponding test sets. Finally, each of these test sets was imputed using both the KNN and SAITS models.

The results presented in Figure \ref{fig:Different_test_drop_rate} confirm that both models are highly effective in recovering photo-$z$ accuracy. When trained on a complete dataset, KNN’s performance is consistently on par with, and marginally better than that of SAITS across all missingness levels. The impact is substantial: even at a high missing fraction of 30\%, KNN imputation reduces the photo-$z$ outlier fraction by a factor of seven compared to the non-imputed baseline.

\subsubsection{Training sample size analysis}\label{Different trainning size}

To evaluate how training sample size influences imputation performance, we trained KNN and SAITS on seven distinct datasets ranging from 5000 to 120 000 samples (see Table \ref{tab:Datasets.}). These models were then evaluated against a fixed test set containing a global missing fraction of 10\%.

The results reveal distinct performance curves for the two models (Figure \ref{fig:different_training_size}). For large training sets ($\geq$ 90 000 samples), both models achieve excellent and comparable performance, with photo-$z$ outlier fractions around 1.5\%. As the training sample size decreases, however, KNN demonstrates superior data efficiency. Its performance degrades gracefully, with the outlier fraction remaining a low 1.7\% even with only 5000 training samples. In contrast, SAITS’s performance degrades more sharply once the training sample size drops below 90 000, reaching an outlier fraction of 2.19\% with a 5000 sample set. This is attributable to the inherent data hunger of DL models; the parameter-heavy Transformer architecture struggles to generalize from small training sets, leading to suboptimal performance compared to the simpler distance-based KNN.

This demonstrates that when a complete training set is available, KNN is not only more accurate but also significantly more data-efficient than SAITS, maintaining high performance even with limited training data.

\begin{figure*}[!t]
    \centering
    \includegraphics[width=1\textwidth]{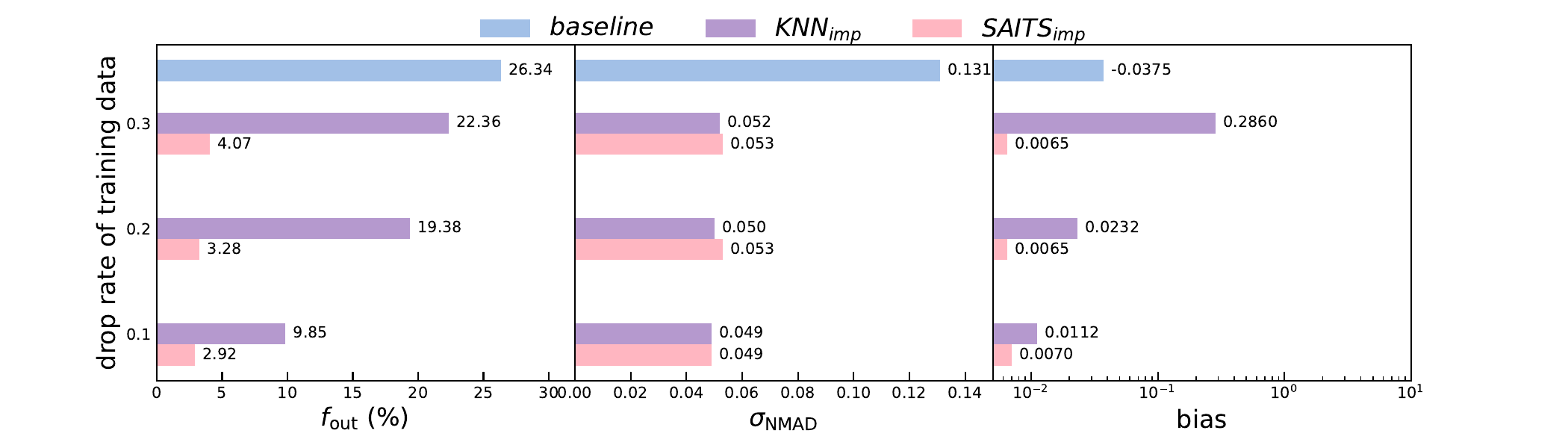}
    \caption{Influence of training sets with different missing rates on model performance. Training sets with missing rates of 10\%, 20\%, and 30\% are used. Blue bars show baseline results from the non-imputed test set (with three bands removed), while pink and purple bars show results after imputation with the KNN and SAITS models, respectively.}
    \label{fig:missing_trainingsets}
\end{figure*}

\begin{figure*}[!t]
    \centering
    \includegraphics[width=1\textwidth]{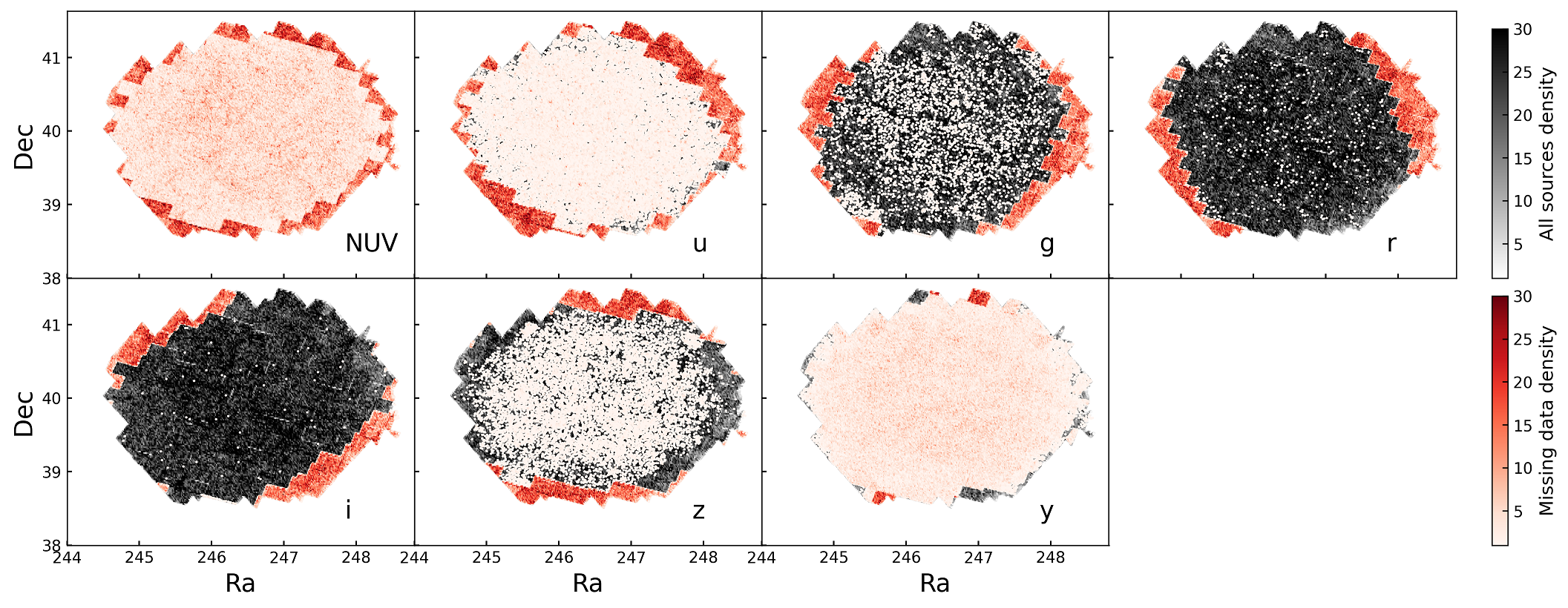}
    \caption{Spatial distribution of missing data for each photometric band in the input catalog. Gray points show all sources, while red points indicate sources with missing data in each band. A higher density of red points corresponds to a higher fraction of missing data.}
    \label{fig:input_location}
\end{figure*}

\subsubsection{Training set with missing data analysis}	\label{The training set contains missing data.}   

A major real-world challenge is the limited availability of complete multi-band training data. When compiling catalogs from multiple surveys, differences in band coverage and survey depth often lead to inconsistent photometric measurements across objects. In addition, flux-limited samples can produce systematic missingness within specific redshift ranges, meaning that restricting the analysis to fully observed sources may introduce selection bias. As a result, constructing a large and unbiased complete training set is not practical. To address this limitation, we evaluate the robustness of our top two models, KNN and SAITS, when trained on incomplete datasets.

We introduced MCAR missingness into the training set at global fractions of 10\%, 20\%, and 30\%. The models were then trained on these incomplete sets and evaluated on the same three bands dropped test set as before. The results, shown in Figure \ref{fig:missing_trainingsets}, reveal a stark difference in robustness. KNN’s performance degrades rapidly as the missing rate of the training set increases, with a 30\% missing rate, its photo-$z$ outlier fraction rises to 22.36\%. In contrast, SAITS demonstrates remarkable stability. Even when trained on data with a 30\% missing rate, its outlier fraction increases only to 4.07\%, which remains approximately six times better than having no imputation at all. These results indicate that KNN requires a complete training set to be effective, whereas SAITS can be reliably trained on highly incomplete data, a significant advantage for practical applications. 

The degradation in KNN performance when applied to incomplete training sets can be attributed to fundamental limitations inherent to distance-based methods. KNN depends critically on accurate distance computations—most commonly Euclidean distance—to identify relevant neighbors. When training data contain missing values, distance calculations are necessarily restricted to the subset of features that are jointly observed, resulting in an effective contraction of the feature space. This contraction often leads to the selection of suboptimal neighbors that fail to reflect the true local structure of the data. In addition, incomplete training sets induce sparsification of the sample space, creating regions where meaningful analogs are absent. As a purely local estimator, KNN lacks mechanisms to compensate for such fragmentation.

In contrast, attention-based models such as SAITS are able to integrate information across features and data points to leverage global dependencies learned through masking-based self-supervision. From a probabilistic perspective, missing data imputation can be interpreted as a conditional inference problem, and the ability of attention mechanisms to selectively aggregate relevant observations may provide a structural advantage in this setting, particularly under irregular or high missingness patterns. Recent theoretical and empirical studies have demonstrated that, in controlled settings with known posteriors, transformer architectures can approximate Bayesian posterior updates with high fidelity \citep{Agarwal2025TheBG}. Although our experiments do not establish that attention-based models perform explicit Bayesian inference, their improved performance is consistent with the view that such architectures are better suited to approximate conditional relationships in incomplete sequential data than distance-based alternatives.

\begin{figure*}[!t]
    \centering	
    \includegraphics[width=1\textwidth]{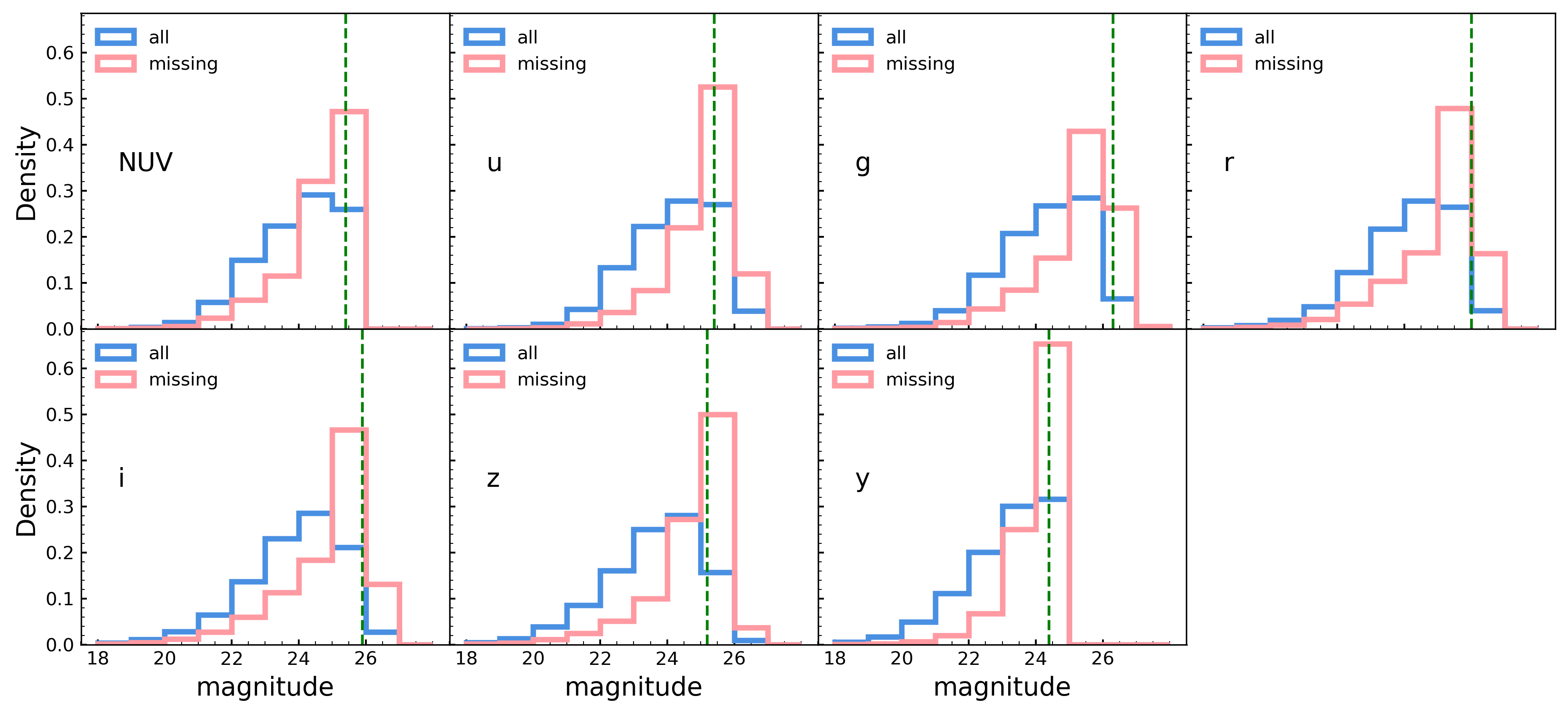}
    \caption{Magnitude distributions of the total source population (blue) and sources with missing data in the output catalog despite being present in the input catalog (pink).The green dashed line shows the magnitude limit of the corresponding CSST band.}
    \label{fig:input_mag}
\end{figure*}

\begin{figure}[!t]
	\centering   
    \includegraphics[width=0.5\textwidth]{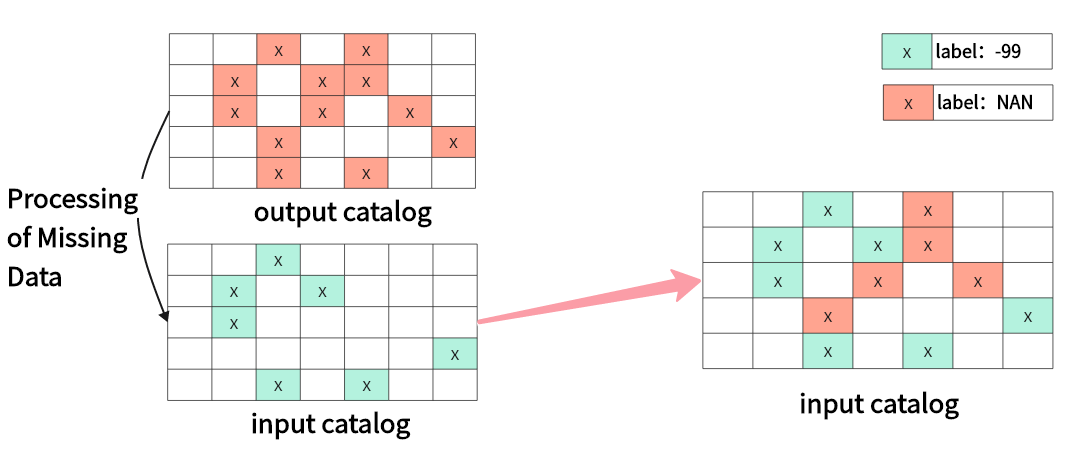}	
    \caption{Schematic illustration of the missing data processing in the input catalog based on the missing pattern of the output catalog.}
	\label{fig:catalog}
\end{figure}

\begin{table}[!t]
	\centering
	\caption{Evaluation metrics for photo-$z$ estimations of the test set with three bands dropped after imputation with models trained on datasets with different missing rates.}  
\fontsize{7pt}{7pt}\selectfont
  \renewcommand{\arraystretch}{1.2}
  \setlength{\tabcolsep}{3.5pt} 
	\begin{tabular}{llll}
        \toprule
		Training set & $f_{out}$(\%)  & $\sigma_{\text{NMAD}}$ & bias\\
		\midrule
		10\% drop rate & 3.12 $\pm$ \num{9.4e-2} 
        & \num{5.5e-02} $\pm$ \num{2.6e-06} & \num{7.5e-03} $\pm$ \num{7.0e-07}  \\
		20\% drop rate & 3.33 $\pm$ \num{5.8e-3} 
        & \num{5.6e-02} $\pm$ \num{1.4e-06} & \num{6.5e-03} $\pm$ \num{5.0e-07}  \\
		30\% drop rate & 3.51 $\pm$ \num{2.4e-1} 
        & \num{5.7e-02} $\pm$ \num{3.9e-06} & \num{7.4e-03} $\pm$ \num{2.6e-06}  \\
		\bottomrule
	\end{tabular}
	\label{tab：The uncertainty of the Saits model}%
\end{table}

\begin{figure*}[!t]
	\centering   
    \includegraphics[width=1\textwidth]{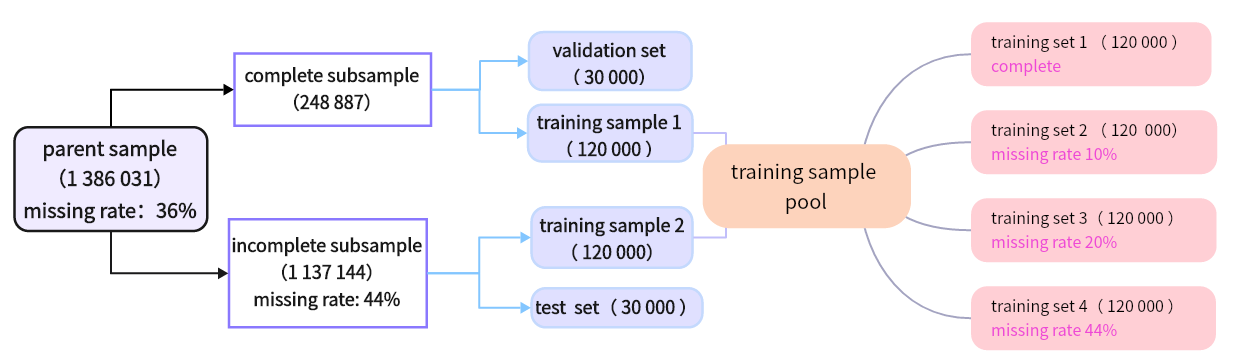}	
    \caption{Dataset partitioning scheme of the CSST mock data. The subsets and their missing rates are shown, with numbers in parentheses indicating the number of sources.}
	\label{fig:datasets}
\end{figure*}

A key difference between the two models is that KNN is deterministic (always producing the same output for a given input), while SAITS is stochastic due to its training process and inherent model architecture. This means each training run of SAITS can produce a slightly different model, leading to variations in the imputed values. We conducted an experiment to quantify this model uncertainty, for each training set (with missing rates of 10\%, 20\%, and 30\%), we trained the SAITS model 10 separate times using the same hyperparameters but different random seeds. Each of the 10 resulting models was then used to impute the same test set (with three bands missing). We then computed the mean and standard deviation of the resulting photo-$z$ metrics across these 10 runs. As shown in Table \ref{tab：The uncertainty of the Saits model}, the model is highly stable. The standard deviations for $\sigma_{\text{NMAD}}$ and bias are exceptionally small (on the order of $10^{-6}$), indicating negligible variation in these metrics. While the standard deviation of the outlier fraction is larger, it remains small, confirming that the imputation results are consistent and reliable across different training runs. This demonstrates that despite its stochastic nature, SAITS provides robust and reproducible performance.

\section{Application on the missing data types of CSST}

\subsection{Missing data properties}

Having established model performance under idealized MCAR conditions, we now turn to more complex and realistic missing data patterns found in simulated CSST catalogs. Such kind of data typically contains a mixture of all three missing types: MCAR, MAR, and MNAR. 

First, the input simulated catalog contains pre-defined gaps. As illustrated by the spatial distribution in Figure \ref{fig:input_location}, this missingness arises from two primary mechanisms. The first is the survey strategy, where edge regions of the footprint lack coverage, resulting in missing data that depends on sky coordinates; this is classified as MAR. The second mechanism involves detection limits, where the simulated observation depth aligns with the survey design. Sources fainter than the magnitude limits are intentionally marked as missing—a classic case of MNAR, as the missingness is inherently dependent on the brightness of the source. This is particularly notable in the $NUV$, $u$, and $y$ bands, which have shallower depths. Overall, the pre-defined missing data in the input catalog is dominated by MNAR mechanism.

Second, additional missingness is introduced in the final output catalog due to simulated observational image detection and pipeline processing. Figure \ref{fig:input_mag} shows the magnitude distribution of sources that were present in the input simulation but were not detected in the output catalog, this missingness occurs for both faint and bright galaxies. Faint galaxies which fall below the signal-to-noise detection threshold are marginally detected, which again constitutes MNAR. For bright sources, however, the reasons for missingness are more complex, including image artifacts, source blending, or other stochastic processing failures. Since this type of missingness is independent of the source’s faintness, it is better described as a combination of MCAR and MAR. Based on an approximate estimate derived from the magnitude limits, MNAR-type missingness accounts for on the order of 30\% of the additional missingness in the output catalog. However, after applying commonly sample selection criteria—retaining only galaxies with detections in more than three photometric bands and with S/N > 10 in the g or r band—this fraction is reduced to about 6\%. Consequently, the additional missingness in the filtered dataset is dominated by MCAR and MAR components, with MNAR playing a relatively minor role.

\subsection{Data preparation}\label{Data preparation}

\begin{figure*}[!t]
	\centering   
    \includegraphics[width=1\textwidth]{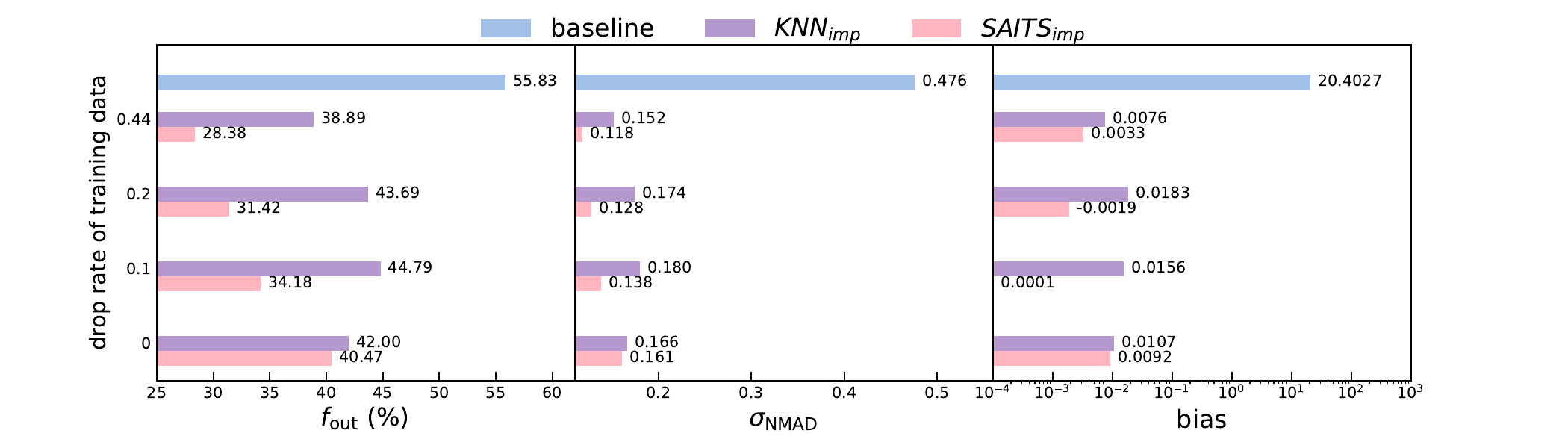}	
    \caption{Effect of training-set missing rates on model performance for realistic CSST missing-pattern data. The missing rates of the training sets are set to 0\%, 10\%, 20\% and 44\%, respectively. Blue bars show the baseline results on non-imputed test set (44\% missing rate), while pink and purple bars display results after KNN and SAITS imputation, respectively.}
	\label{fig:CSST_imp}
\end{figure*} 

The dataset used here is no longer a complete subsample of the input catalog, but instead consists of all galaxies from the input catalog with high quality output photometry and detections in at least two bands. The final dataset contains 1 386 031 galaxies with an overall missing data fraction of 36\%. This fraction rises to 44\% when considering only the subset of galaxies with missing data. To create a realistic testbed for our imputation models, one that mirrors the complex missing data patterns of the final CSST survey while avoiding artifacts from the simulation pipeline, we devised a specific data preparation strategy. The core of this strategy is to map the missingness pattern from the output catalog onto the pristine input catalog. This approach allows us to work with the ground-truth photometry of the input simulation while testing the models against a realistic, multi-mechanism missing data pattern. The procedure is as follows:

\begin{enumerate}
    \item Map missingness pattern: the missingness pattern from the output catalog is mapped onto the input catalog (Figure \ref{fig:catalog}). This process generates two distinct flags for missing values. Values originally missing in the input catalog are marked as -99; these are primarily systematic MNAR cases resulting from magnitude limit cuts. Values present in the input catalog but missing in the output are marked as NAN; these represent the more stochastic MCAR/MAR cases arising from observational simulation and pipeline processing.
    \item Partition dataset: the dataset is partitioned according to the logic illustrated in Figure \ref{fig:datasets}. First, we divide the entire dataset into a complete subset (no missing values) and an incomplete subset (at least one band missing,  the overall missing rate is 44\%). The validation set is drawn exclusively from the complete subset, while the test set is drawn exclusively from the incomplete subset. The remaining data are combined to form a pool for constructing the training sets with different missing rates.
    \item Training data sampling: constructed four distinct training sets from the training pool for our experiments: (a) A complete training set derived from the remaining complete data, and (b) Three incomplete training sets with overall missing fractions of 10\%, 20\%, and 44\% by mixing the remaining complete and incomplete dataset with different ratios.
\end{enumerate}

\begin{table}[!t]
	\centering
	\caption{Photo-$z$ metrics for sources with different missing type. 
    }
	\resizebox{\linewidth}{!}{
    \begin{tabular}{lcc}
		\toprule
	Missing type   & Non-imputed  & After imputation  \\	
    & $f_{\text{out}}(\%)/\sigma_{\text{NMAD}}/\text{bias}$ & $f_{\text{out}}(\%)/\sigma_{\text{NMAD}}/\text{bias}$\\
		\midrule
		Only NAN & 7.30/0.084/-0.0795 & 3.57/0.063/-0.0038 \\
		Only -99 & 2.57/0.085/-0.0078  & 2.67/0.095/0.0552 \\
		Mix & 5.74/0.132/-0.0765 & 3.77/0.093/0.0235 \\
		\midrule
		all & 15.61/0.100/-0.0645 & 10.01/0.073/0.0247\\
		\bottomrule
	\end{tabular}}
    \tablefoot{The test set is restricted to galaxies with detections in $>3$ photometric bands and $g$- or $r$-band S/N $>10$. The imputation model is SAITS.}
	\label{tab:compare_three_type} 
\end{table}

\begin{figure*}[!t]
	\centering   
    \includegraphics[width=1\textwidth]{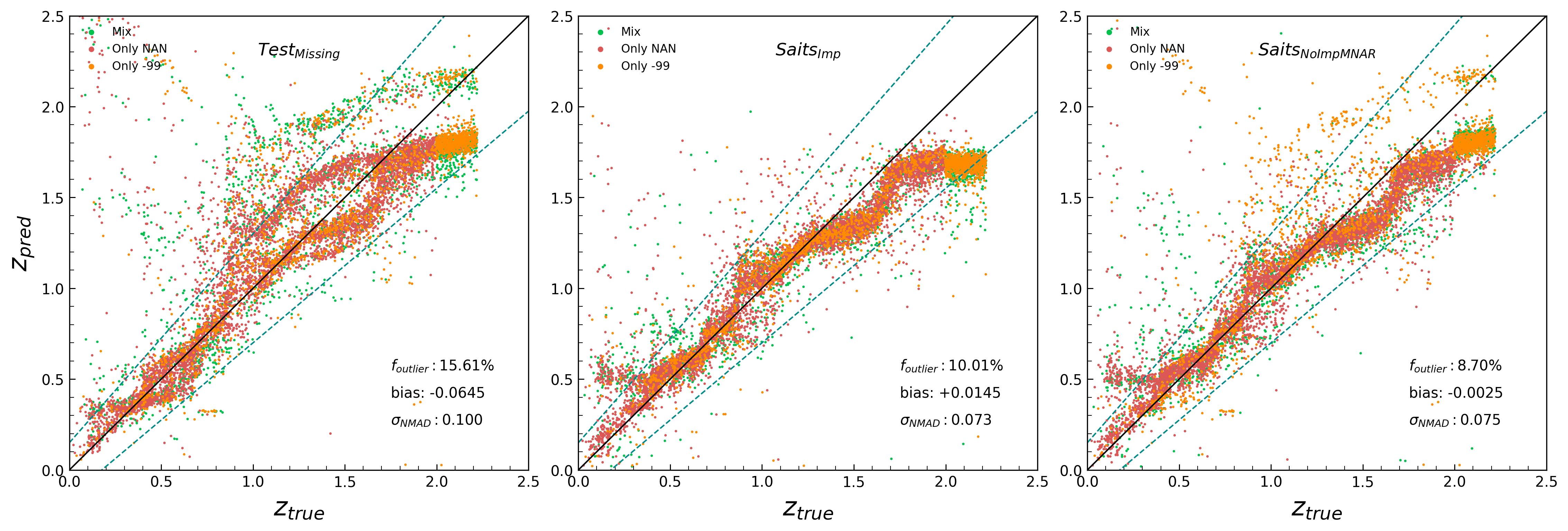}	
    \caption{Photo-$z$ versus true values for test set galaxies with detections in $>3$ photometric bands and $g$ or $r$ band S/N $>10$. The left panel shows results for the non-imputed test set. The middle panel shows results after SAITS imputation. The right panel shows results where “NAN” and "Mix" missing types are imputed with SAITS, while values flagged as “–99” are not imputed. }
	\label{fig:multiplot_csst}
\end{figure*}

\begin{table*}[!t]
	\centering
	\caption{Photo-$z$ metrics for test-set galaxies missing 1–3 bands, with detections in more than three photometric bands and $g$- or $r$-band S/N $>10$.}
    \setlength{\tabcolsep}{23pt}
    \begin{tabular}{lccc}
		\toprule
	  Source type & Non-imputed & Full imputed & MAR/MCAR imputed\\	
        & $f_{\text{out}}(\%)/\sigma_{\text{NMAD}}/\text{bias}$ 
        & $f_{\text{out}}(\%)/\sigma_{\text{NMAD}}/\text{bias}$ 
        & $f_{\text{out}}(\%)/\sigma_{\text{NMAD}}/\text{bias}$ \\
		\midrule
		1 missing band  & 5.34/0.075/-0.0156 & 6.26/0.065/0.0257 & 2.52/0.064/0.0199 \\
		2 missing bands  & 15.74/0.104/-0.0546 & 10.29/0.077/0.0152 & 7.78/0.077/0.0025 \\
		3 missing bands  & 35.38/0.174/-0.1737 & 16.92/0.091/-0.0083 & 22.04/0.110/-0.0532 \\
		\bottomrule
	\end{tabular}
    \tablefoot{Results are shown for different imputation strategies and compared to the pre-imputation baseline. The imputation model is SAITS.}
    \label{tab:compare_different_missingband} 
\end{table*}

\subsection{Results analysis}

In this section, we evaluate the performance of the imputation models on a fixed test set with an overall missing data fraction of 44\%, consistent with the estimated missing rate from the CSST observational simulation. To investigate how the missing rate of the training dataset affects the model performance, we train both the KNN and SAITS models on four distinct training datasets described in Section \ref{Data preparation}.

The results, presented in Figure \ref{fig:CSST_imp}, reveal an crucial insight that differs from the trend observed in our idealized experiments in Section \ref{Model imputation capability experiments}. In those experiments, where the datasets contain only MCAR-type missing values, lower missing rates consistently lead to higher photo-$z$ accuracy  (Figure \ref{fig:missing_trainingsets}). In contrast, under the more realistic setting considered here, where the data exhibit a mixture of MCAR, MAR, and MNAR missing patterns, training on incomplete data yields significantly better imputation performance than training on fully complete datasets. The optimal performance is achieved when the training set’s missing fraction matches that of the test set (44\%). In this scenario, SAITS reduces the photo-$z$ outlier fraction by nearly half, from a baseline of 55.83\% for the non-imputed data down to 28.38\%, and consistently outperforms KNN. These results demonstrate that, for real-world applications, training an imputation model on data that statistically mirrors the target dataset, in terms of both the quantity and the pattern of missingness, is crucial for achieving optimal performance.

To investigate the imputation performance on different missing data patterns, we first filtered the test set to include only galaxies with detections in more than three bands photometry and S/N $>10$ in the $g$ or $r$ band, and classified the subset into three mutually exclusive groups based on their missing data flags described in Section \ref{Data preparation}:

\begin{itemize}
    \item MCAR/MAR group (Only NAN flag): galaxies where all missing bands were pipeline-induced.
    \item MNAR group (Only -99 flag): galaxies where all missing bands were pre-defined by the magnitude limit.
    \item Mix group: galaxies that contain both missing types.
\end{itemize}

The results of the best-performing SAITS model, summarized in Table \ref{tab:compare_three_type}, are notable. For the MCAR/MAR group, the imputation is highly effective, reducing the outlier fraction by a factor of two compared to the baseline. Conversely, for the MNAR group, imputation is detrimental, slightly degrading the photo-$z$ accuracy to a level worse than the non-imputed data. Because a large fraction of missingness is driven by the survey magnitude limit, the overall outlier fraction decreases by only 5.6\%. 

This finding suggests a clear optimal strategy: selectively impute only the data known to be MCAR or MAR. We tested this hybrid strategy by imputing only the NAN-flagged values, while leaving the -99 values missing. As shown in Figure \ref{fig:multiplot_csst}, this method yields the best overall performance, reducing the outlier fraction for these sources from 10.01\% (with full imputation) to a superior 8.7\%. This improvement arises because MNAR values, which are often driven by magnitude limits, tend to be imputed with artificially brighter fluxes, leading to systematically underestimated redshifts. This effect is visible at the high-redshift end of the middle panel in Figure \ref{fig:multiplot_csst}, where missing measurements are more likely driven by the depth of the survey rather than random absence. These results demonstrate that although imputation is effective for MCAR/MAR missingness, it is preferable to avoid imputing values that are missing due to physical limits (MNAR), where the absence of information is itself informative for photo-$z$ estimation. 

For further investigation, we group galaxies by the number of missing bands and compute photo-$z$ metrics for each subset separately. The results, presented in Table \ref{tab:compare_different_missingband}, show that galaxies with three missing bands exhibit substantially degraded photo-$z$ performance in both the non-imputed and imputed cases. Therefore, for statistical analysis requiring high photo-$z$ accuracy, it is conservative to restrict the sample to galaxies with fewer than three missing bands.

\section{Conclusions}

This study presents a systematic evaluation of data imputation methods to improve photo-$z$ estimation accuracy. Our investigation began by assessing a suite of ML and DL models to identify the top performers. We then subjected these leading models, KNN and SAITS, to comprehensive robustness tests, varying both the training sample size and the missing data rate. Finally, we evaluated their performance on a realistic testbed designed to represent the complex, mixed-mechanism missing data pattern of the CSST survey, providing a practical validation under real observational conditions.

Under idealized MCAR conditions with a complete training set, the deterministic KNN model provides the best overall imputation performance, closely followed by the attention-based SAITS model. Both models are remarkably data-efficient, maintaining high accuracy even with training datasets as small as 5000 samples. This specific finding reflects a broader trend among the different architectural foundations tested: among ML category, distance-based models like KNN outperformed tree-based models, while in DL approaches, transformer-based models like SAITS were superior.

The robustness of the models to incomplete training data differs starkly. KNN’s performance degrades sharply as the missing fraction of the training set increases. In contrast, SAITS demonstrates exceptional robustness, maintaining high performance even when trained on highly incomplete data.

When tested on the realistic, mixed-mechanism CSST missing data pattern, the performance ranking differs: SAITS consistently outperforms KNN, optimal results are achieved when the missing data pattern of the training set statistically mirrors that of the test set. This highlights a crucial principle for any supervised learning approach: maintaining domain consistency between training and testing samples, where the distributions of both features and targets are aligned, is critical for regression-based tasks such as imputation.

Although creating domain-consistent datasets is relatively straightforward for an imputation task, it is considerably more challenging for many other regression problems in astronomy. Practical observational datasets are often highly imbalanced and subject to complex selection biases, which must be carefully assessed when deploying supervised learning techniques. Consequently, a model trained and validated on a domain-consistent dataset is likely to experience performance degradation when applied to real-world scenarios where a "domain shift” exists. Researchers must be cautious about this issue and consider strategies such as domain adaptation to address its impact. Otherwise, accuracy assessments derived from idealized test datasets may not serve as reliable indicators of real-world performance.

For the mixed types of missing data common in practical observations, we find that while general imputation models like SAITS are highly effective for MCAR and MAR data, they are detrimental for MNAR data. In multi-band photometry, MNAR data primarily arise from magnitude-limited observations, meaning the missingness itself contains physical information. Statistical models like SAITS are designed to minimize the reconstruction error based on the available data distribution, which is often biased toward brighter sources in supervised training sets.

A straightforward and effective strategy is therefore to impute only MAR and MCAR data, while leaving MNAR values untouched. Implementing this requires forced photometry that can distinguish between stochastic missingness (MCAR/MAR) and non-detections (MNAR). We also attempted to teach the model the MNAR pattern by training on deep-field data, but this strategy was ineffective, demonstrating that the MNAR pattern is not a purely statistical phenomenon that can be learned by proxy. This inherent characteristic of astronomical data is difficult to incorporate as a constraint into general-purpose imputation models. We therefore recommend the development of specialized architectures capable of disentangling these distinct missingness mechanisms. Such models should treat stochastic gaps (MCAR/MAR) through traditional reconstruction while incorporating magnitude limits as informative priors for MNAR, thereby preserving the physical integrity of the completed catalog. 

\begin{acknowledgements}
      This work is supported by the China Manned Space Project with Grant NO. CMS-CSST-2021-A07, No. CMS-CSST-2021-A01, and NO. CMS-CSST-2025-A05, ZC and LPF acknowledge support from NSFC grant No. 12541302, No. 12141302, and the Innovation Program of Shanghai Municipal Education Commission (Grant No. 2025GDZKZD04).
\end{acknowledgements}

\bibliographystyle{aa} 
\bibliography{aa.bib}

@article{Brammer_2008,
doi = {10.1086/591786},
url = {https://dx.doi.org/10.1086/591786},
year = {2008},
month = {oct},
publisher = {},
volume = {686},
number = {2},
pages = {1503},
author = {Brammer, Gabriel B. and van Dokkum, Pieter G. and Coppi, Paolo},
title = {EAZY: A Fast, Public Photometric Redshift Code},
journal = {\apj},
}

@article{du2023pypots,
           author = {{Du}, Wenjie and {Yang}, Yiyuan and {Qian}, Linglong and {Wang}, Jun and {Wen}, Qingsong},
        title = "{PyPOTS: A Python Toolkit for Machine Learning on Partially-Observed Time Series}",
      journal = {arXiv e-prints},
         year = 2023,
        month = may,
          doi = {10.48550/arXiv.2305.18811},
archivePrefix = {arXiv},
       eprint = {2305.18811},
 primaryClass = {stat.ML},
       adsurl = {https://ui.adsabs.harvard.edu/abs/2023arXiv230518811D}
}

@ARTICLE{knn,
  author={Cover, T. and Hart, P.},
  journal={IEEE Transactions on Information Theory}, 
  title={Nearest neighbor pattern classification}, 
  year={1967},
  volume={13},
  number={1},
  pages={21-27},
  doi={10.1109/TIT.1967.1053964}}

@inproceedings{2018CatBoost,
author = {Prokhorenkova, Liudmila and Gusev, Gleb and Vorobev, Aleksandr and Dorogush, Anna Veronika and Gulin, Andrey},
title = {CatBoost: unbiased boosting with categorical features},
year = {2018},
publisher = {Curran Associates Inc.},
address = {Red Hook, NY, USA},
booktitle = {Proceedings of the 32nd International Conference on Neural Information Processing Systems},
pages = {6639–6649},
numpages = {11},
location = {Montr\'{e}al, Canada},
series = {NIPS'18}
}

@article{2001RF,
  title={Random forests, machine learning 45},
  author={ Breiman, L. },
  journal={Journal of Clinical Microbiology},
  volume={2},
  pages={199-228},
  year={2001},
}

@inproceedings{Transformer,
        author = {Vaswani, Ashish and Shazeer, Noam and Parmar, Niki and Uszkoreit, Jakob and Jones, Llion and Gomez, Aidan N and Kaiser, \L ukasz and Polosukhin, Illia},
 booktitle = {Advances in Neural Information Processing Systems},
 editor = {I. Guyon and U. Von Luxburg and S. Bengio and H. Wallach and R. Fergus and S. Vishwanathan and R. Garnett},
 pages = {},
 publisher = {Curran Associates, Inc.},
 title = {Attention is All you Need},
 url = {https://proceedings.neurips.cc/paper_files/paper/2017/file/3f5ee243547dee91fbd053c1c4a845aa-Paper.pdf},
 volume = {30},
 year = {2017}
}

@inproceedings{itransformer,
title={iTransformer: Inverted Transformers Are Effective for Time Series Forecasting},
author={Yong Liu and Tengge Hu and Haoran Zhang and Haixu Wu and Shiyu Wang and Lintao Ma and Mingsheng Long},
booktitle={The Twelfth International Conference on Learning Representations},
year={2024},
url={https://openreview.net/forum?id=JePfAI8fah}
}

@article{SAITS,
title = {SAITS: Self-attention-based imputation for time series},
journal = {Expert Systems with Applications},
volume = {219},
pages = {119619},
year = {2023},
issn = {0957-4174},
doi = {https://doi.org/10.1016/j.eswa.2023.119619},
url = {https://www.sciencedirect.com/science/article/pii/S0957417423001203},
author = {Wenjie Du and David Côté and Yan Liu},
}

@ARTICLE{MRNN,
  author={Yoon, Jinsung and Zame, William R. and van der Schaar, Mihaela},
  journal={IEEE Transactions on Biomedical Engineering}, 
  title={Estimating Missing Data in Temporal Data Streams Using Multi-Directional Recurrent Neural Networks}, 
  year={2019},
  volume={66},
  number={5},
  pages={1477-1490},
  doi={10.1109/TBME.2018.2874712}}

@inproceedings{BRITs,
        author = {Cao, Wei and Wang, Dong and Li, Jian and Zhou, Hao and Li, Lei and Li, Yitan},
 booktitle = {Advances in Neural Information Processing Systems},
 pages = {},
 publisher = {Curran Associates, Inc.},
 title = {BRITS: Bidirectional Recurrent Imputation for Time Series},
 url = {https://proceedings.neurips.cc/paper_files/paper/2018/file/734e6bfcd358e25ac1db0a4241b95651-Paper.pdf},
 volume = {31},
 year = {2018}
}

@InProceedings{GP-VAE,
  title = 	 {GP-VAE: Deep Probabilistic Time Series Imputation},
  author =       {Fortuin, Vincent and Baranchuk, Dmitry and Raetsch, Gunnar and Mandt, Stephan},
  booktitle = 	 {Proceedings of the Twenty Third International Conference on Artificial Intelligence and Statistics},
  pages = 	 {1651--1661},
  year = 	 {2020},
  editor = 	 {Chiappa, Silvia and Calandra, Roberto},
  volume = 	 {108},
  series = 	 {Proceedings of Machine Learning Research},
  month = 	 {26--28 Aug},
  publisher =    {PMLR},
  url = 	 {https://proceedings.mlr.press/v108/fortuin20a.html},
}

@article{usgan, 
title={Generative Semi-supervised Learning for Multivariate Time Series Imputation}, volume={35}, url={https://ojs.aaai.org/index.php/AAAI/article/view/17086}, DOI={10.1609/aaai.v35i10.17086}, abstractNote={The missing values, widely existed in multivariate time series data, hinder the effective data analysis. Existing time series imputation methods do not make full use of the label information in real-life time series data. In this paper, we propose a novel semi-supervised generative adversarial network model, named SSGAN, for missing value imputation in multivariate time series data. It consists of three players, i.e., a generator, a discriminator, and a classifier. The classifier predicts labels of time series data, and thus it drives the generator to estimate the missing values (or components), conditioned on observed components and data labels at the same time. We introduce a temporal reminder matrix to help the discriminator better distinguish the observed components from the imputed ones. Moreover, we theoretically prove that, SSGAN using the temporal reminder matrix and the classifier does learn to estimate missing values converging to the true data distribution when the Nash equilibrium is achieved. Extensive experiments on three public real-world datasets demonstrate that, SSGAN yields a more than 15% gain in performance, compared with the state-of-the-art methods.}, number={10}, journal={Proceedings of the AAAI Conference on Artificial Intelligence}, author={Miao, Xiaoye and Wu, Yangyang and Wang, Jun and Gao, Yunjun and Mao, Xudong and Yin, Jianwei}, year={2021}, month={May}, pages={8983-8991} }

@ARTICLE{2011Euclid,
       author = {{Laureijs}, R. and {Amiaux}, J. and {Arduini}, S. and {Augu{\`e}res}, J. -L. and {Brinchmann}, J. and {Cole}, R. and {Cropper}, M. and {Dabin}, C. and {Duvet}, L. and {Ealet}, A. and {Garilli}, B. and {Gondoin}, P. and {Guzzo}, L. and {Hoar}, J. and {Hoekstra}, H. and {Holmes}, R. and {Kitching}, T. and {Maciaszek}, T. and {Mellier}, Y. and {Pasian}, F. and {Percival}, W. and {Rhodes}, J. and {Saavedra Criado}, G. and {Sauvage}, M. and {Scaramella}, R. and {Valenziano}, L. and {Warren}, S. and {Bender}, R. and {Castander}, F. and {Cimatti}, A. and {Le F{\`e}vre}, O. and {Kurki-Suonio}, H. and {Levi}, M. and {Lilje}, P. and {Meylan}, G. and {Nichol}, R. and {Pedersen}, K. and {Popa}, V. and {Rebolo Lopez}, R. and {Rix}, H. -W. and {Rottgering}, H. and {Zeilinger}, W. and {Grupp}, F. and {Hudelot}, P. and {Massey}, R. and {Meneghetti}, M. and {Miller}, L. and {Paltani}, S. and {Paulin-Henriksson}, S. and {Pires}, S. and {Saxton}, C. and {Schrabback}, T. and {Seidel}, G. and {Walsh}, J. and {Aghanim}, N. and {Amendola}, L. and {Bartlett}, J. and {Baccigalupi}, C. and {Beaulieu}, J. -P. and {Benabed}, K. and {Cuby}, J. -G. and {Elbaz}, D. and {Fosalba}, P. and {Gavazzi}, G. and {Helmi}, A. and {Hook}, I. and {Irwin}, M. and {Kneib}, J. -P. and {Kunz}, M. and {Mannucci}, F. and {Moscardini}, L. and {Tao}, C. and {Teyssier}, R. and {Weller}, J. and {Zamorani}, G. and {Zapatero Osorio}, M.~R. and {Boulade}, O. and {Foumond}, J.~J. and {Di Giorgio}, A. and {Guttridge}, P. and {James}, A. and {Kemp}, M. and {Martignac}, J. and {Spencer}, A. and {Walton}, D. and {Bl{\"u}mchen}, T. and {Bonoli}, C. and {Bortoletto}, F. and {Cerna}, C. and {Corcione}, L. and {Fabron}, C. and {Jahnke}, K. and {Ligori}, S. and {Madrid}, F. and {Martin}, L. and {Morgante}, G. and {Pamplona}, T. and {Prieto}, E. and {Riva}, M. and {Toledo}, R. and {Trifoglio}, M. and {Zerbi}, F. and {Abdalla}, F. and {Douspis}, M. and {Grenet}, C. and {Borgani}, S. and {Bouwens}, R. and {Courbin}, F. and {Delouis}, J. -M. and {Dubath}, P. and {Fontana}, A. and {Frailis}, M. and {Grazian}, A. and {Koppenh{\"o}fer}, J. and {Mansutti}, O. and {Melchior}, M. and {Mignoli}, M. and {Mohr}, J. and {Neissner}, C. and {Noddle}, K. and {Poncet}, M. and {Scodeggio}, M. and {Serrano}, S. and {Shane}, N. and {Starck}, J. -L. and {Surace}, C. and {Taylor}, A. and {Verdoes-Kleijn}, G. and {Vuerli}, C. and {Williams}, O.~R. and {Zacchei}, A. and {Altieri}, B. and {Escudero Sanz}, I. and {Kohley}, R. and {Oosterbroek}, T. and {Astier}, P. and {Bacon}, D. and {Bardelli}, S. and {Baugh}, C. and {Bellagamba}, F. and {Benoist}, C. and {Bianchi}, D. and {Biviano}, A. and {Branchini}, E. and {Carbone}, C. and {Cardone}, V. and {Clements}, D. and {Colombi}, S. and {Conselice}, C. and {Cresci}, G. and {Deacon}, N. and {Dunlop}, J. and {Fedeli}, C. and {Fontanot}, F. and {Franzetti}, P. and {Giocoli}, C. and {Garcia-Bellido}, J. and {Gow}, J. and {Heavens}, A. and {Hewett}, P. and {Heymans}, C. and {Holland}, A. and {Huang}, Z. and {Ilbert}, O. and {Joachimi}, B. and {Jennins}, E. and {Kerins}, E. and {Kiessling}, A. and {Kirk}, D. and {Kotak}, R. and {Krause}, O. and {Lahav}, O. and {van Leeuwen}, F. and {Lesgourgues}, J. and {Lombardi}, M. and {Magliocchetti}, M. and {Maguire}, K. and {Majerotto}, E. and {Maoli}, R. and {Marulli}, F. and {Maurogordato}, S. and {McCracken}, H. and {McLure}, R. and {Melchiorri}, A. and {Merson}, A. and {Moresco}, M. and {Nonino}, M. and {Norberg}, P. and {Peacock}, J. and {Pello}, R. and {Penny}, M. and {Pettorino}, V. and {Di Porto}, C. and {Pozzetti}, L. and {Quercellini}, C. and {Radovich}, M. and {Rassat}, A. and {Roche}, N. and {Ronayette}, S. and {Rossetti}, E.},
        title = "{Euclid Definition Study Report}",
      journal = {arXiv e-prints},
         year = 2011,
        month = oct,
          doi = {10.48550/arXiv.1110.3193},
archivePrefix = {arXiv},
       eprint = {1110.3193},
 primaryClass = {astro-ph.CO},
       adsurl = {https://ui.adsabs.harvard.edu/abs/2011arXiv1110.3193L}
}

@ARTICLE{2009LSST,
       author = {{LSST Science Collaboration} and {Abell}, Paul A. and {Allison}, Julius and {Anderson}, Scott F. and {Andrew}, John R. and {Angel}, J. Roger P. and {Armus}, Lee and {Arnett}, David and {Asztalos}, S.~J. and {Axelrod}, Tim S. and {Bailey}, Stephen and {Ballantyne}, D.~R. and {Bankert}, Justin R. and {Barkhouse}, Wayne A. and {Barr}, Jeffrey D. and {Barrientos}, L. Felipe and {Barth}, Aaron J. and {Bartlett}, James G. and {Becker}, Andrew C. and {Becla}, Jacek and {Beers}, Timothy C. and {Bernstein}, Joseph P. and {Biswas}, Rahul and {Blanton}, Michael R. and {Bloom}, Joshua S. and {Bochanski}, John J. and {Boeshaar}, Pat and {Borne}, Kirk D. and {Bradac}, Marusa and {Brandt}, W.~N. and {Bridge}, Carrie R. and {Brown}, Michael E. and {Brunner}, Robert J. and {Bullock}, James S. and {Burgasser}, Adam J. and {Burge}, James H. and {Burke}, David L. and {Cargile}, Phillip A. and {Chandrasekharan}, Srinivasan and {Chartas}, George and {Chesley}, Steven R. and {Chu}, You-Hua and {Cinabro}, David and {Claire}, Mark W. and {Claver}, Charles F. and {Clowe}, Douglas and {Connolly}, A.~J. and {Cook}, Kem H. and {Cooke}, Jeff and {Cooray}, Asantha and {Covey}, Kevin R. and {Culliton}, Christopher S. and {de Jong}, Roelof and {de Vries}, Willem H. and {Debattista}, Victor P. and {Delgado}, Francisco and {Dell'Antonio}, Ian P. and {Dhital}, Saurav and {Di Stefano}, Rosanne and {Dickinson}, Mark and {Dilday}, Benjamin and {Djorgovski}, S.~G. and {Dobler}, Gregory and {Donalek}, Ciro and {Dubois-Felsmann}, Gregory and {Durech}, Josef and {Eliasdottir}, Ardis and {Eracleous}, Michael and {Eyer}, Laurent and {Falco}, Emilio E. and {Fan}, Xiaohui and {Fassnacht}, Christopher D. and {Ferguson}, Harry C. and {Fernandez}, Yanga R. and {Fields}, Brian D. and {Finkbeiner}, Douglas and {Figueroa}, Eduardo E. and {Fox}, Derek B. and {Francke}, Harold and {Frank}, James S. and {Frieman}, Josh and {Fromenteau}, Sebastien and {Furqan}, Muhammad and {Galaz}, Gaspar and {Gal-Yam}, A. and {Garnavich}, Peter and {Gawiser}, Eric and {Geary}, John and {Gee}, Perry and {Gibson}, Robert R. and {Gilmore}, Kirk and {Grace}, Emily A. and {Green}, Richard F. and {Gressler}, William J. and {Grillmair}, Carl J. and {Habib}, Salman and {Haggerty}, J.~S. and {Hamuy}, Mario and {Harris}, Alan W. and {Hawley}, Suzanne L. and {Heavens}, Alan F. and {Hebb}, Leslie and {Henry}, Todd J. and {Hileman}, Edward and {Hilton}, Eric J. and {Hoadley}, Keri and {Holberg}, J.~B. and {Holman}, Matt J. and {Howell}, Steve B. and {Infante}, Leopoldo and {Ivezic}, Zeljko and {Jacoby}, Suzanne H. and {Jain}, Bhuvnesh and {R} and {Jedicke} and {Jee}, M. James and {Garrett Jernigan}, J. and {Jha}, Saurabh W. and {Johnston}, Kathryn V. and {Jones}, R. Lynne and {Juric}, Mario and {Kaasalainen}, Mikko and {Styliani} and {Kafka} and {Kahn}, Steven M. and {Kaib}, Nathan A. and {Kalirai}, Jason and {Kantor}, Jeff and {Kasliwal}, Mansi M. and {Keeton}, Charles R. and {Kessler}, Richard and {Knezevic}, Zoran and {Kowalski}, Adam and {Krabbendam}, Victor L. and {Krughoff}, K. Simon and {Kulkarni}, Shrinivas and {Kuhlman}, Stephen and {Lacy}, Mark and {Lepine}, Sebastien and {Liang}, Ming and {Lien}, Amy and {Lira}, Paulina and {Long}, Knox S. and {Lorenz}, Suzanne and {Lotz}, Jennifer M. and {Lupton}, R.~H. and {Lutz}, Julie and {Macri}, Lucas M. and {Mahabal}, Ashish A. and {Mandelbaum}, Rachel and {Marshall}, Phil and {May}, Morgan and {McGehee}, Peregrine M. and {Meadows}, Brian T. and {Meert}, Alan and {Milani}, Andrea and {Miller}, Christopher J. and {Miller}, Michelle and {Mills}, David and {Minniti}, Dante and {Monet}, David and {Mukadam}, Anjum S. and {Nakar}, Ehud and {Neill}, Douglas R. and {Newman}, Jeffrey A. and {Nikolaev}, Sergei and {Nordby}, Martin and {O'Connor}, Paul and {Oguri}, Masamune and {Oliver}, John and {Olivier}, Scot S. and {Olsen}, Julia K. and {Olsen}, Knut and {Olszewski}, Edward W. and {Oluseyi}, Hakeem and {Padilla}, Nelson D. and {Parker}, Alex and {Pepper}, Joshua and {Peterson}, John R. and {Petry}, Catherine and {Pinto}, Philip A. and {Pizagno}, James L. and {Popescu}, Bogdan and {Prsa}, Andrej and {Radcka}, Veljko and {Raddick}, M. Jordan and {Rasmussen}, Andrew and {Rau}, Arne and {Rho}, Jeonghee and {Rhoads}, James E. and {Richards}, Gordon T. and {Ridgway}, Stephen T. and {Robertson}, Brant E. and {Roskar}, Rok and {Saha}, Abhijit and {Sarajedini}, Ata and {Scannapieco}, Evan and {Schalk}, Terry and {Schindler}, Rafe and {Schmidt}, Samuel},
        title = "{LSST Science Book, Version 2.0}",
      journal = {arXiv e-prints},
         year = 2009,
        month = dec,
          doi = {10.48550/arXiv.0912.0201},
archivePrefix = {arXiv},
       eprint = {0912.0201},
 primaryClass = {astro-ph.IM},
       adsurl = {https://ui.adsabs.harvard.edu/abs/2009arXiv0912.0201L}
}

@ARTICLE{2019LSST,
       author = {{Ivezi{\'c}}, {\v{Z}}eljko and {Kahn}, Steven M. and {Tyson}, J. Anthony and {Abel}, Bob and {Acosta}, Emily and {Allsman}, Robyn and {Alonso}, David and {AlSayyad}, Yusra and {Anderson}, Scott F. and {Andrew}, John and {Angel}, James Roger P. and {Angeli}, George Z. and {Ansari}, Reza and {Antilogus}, Pierre and {Araujo}, Constanza and {Armstrong}, Robert and {Arndt}, Kirk T. and {Astier}, Pierre and {Aubourg}, {\'E}ric and {Auza}, Nicole and {Axelrod}, Tim S. and {Bard}, Deborah J. and {Barr}, Jeff D. and {Barrau}, Aurelian and {Bartlett}, James G. and {Bauer}, Amanda E. and {Bauman}, Brian J. and {Baumont}, Sylvain and {Bechtol}, Ellen and {Bechtol}, Keith and {Becker}, Andrew C. and {Becla}, Jacek and {Beldica}, Cristina and {Bellavia}, Steve and {Bianco}, Federica B. and {Biswas}, Rahul and {Blanc}, Guillaume and {Blazek}, Jonathan and {Blandford}, Roger D. and {Bloom}, Josh S. and {Bogart}, Joanne and {Bond}, Tim W. and {Booth}, Michael T. and {Borgland}, Anders W. and {Borne}, Kirk and {Bosch}, James F. and {Boutigny}, Dominique and {Brackett}, Craig A. and {Bradshaw}, Andrew and {Brandt}, William Nielsen and {Brown}, Michael E. and {Bullock}, James S. and {Burchat}, Patricia and {Burke}, David L. and {Cagnoli}, Gianpietro and {Calabrese}, Daniel and {Callahan}, Shawn and {Callen}, Alice L. and {Carlin}, Jeffrey L. and {Carlson}, Erin L. and {Chandrasekharan}, Srinivasan and {Charles-Emerson}, Glenaver and {Chesley}, Steve and {Cheu}, Elliott C. and {Chiang}, Hsin-Fang and {Chiang}, James and {Chirino}, Carol and {Chow}, Derek and {Ciardi}, David R. and {Claver}, Charles F. and {Cohen-Tanugi}, Johann and {Cockrum}, Joseph J. and {Coles}, Rebecca and {Connolly}, Andrew J. and {Cook}, Kem H. and {Cooray}, Asantha and {Covey}, Kevin R. and {Cribbs}, Chris and {Cui}, Wei and {Cutri}, Roc and {Daly}, Philip N. and {Daniel}, Scott F. and {Daruich}, Felipe and {Daubard}, Guillaume and {Daues}, Greg and {Dawson}, William and {Delgado}, Francisco and {Dellapenna}, Alfred and {de Peyster}, Robert and {de Val-Borro}, Miguel and {Digel}, Seth W. and {Doherty}, Peter and {Dubois}, Richard and {Dubois-Felsmann}, Gregory P. and {Durech}, Josef and {Economou}, Frossie and {Eifler}, Tim and {Eracleous}, Michael and {Emmons}, Benjamin L. and {Fausti Neto}, Angelo and {Ferguson}, Henry and {Figueroa}, Enrique and {Fisher-Levine}, Merlin and {Focke}, Warren and {Foss}, Michael D. and {Frank}, James and {Freemon}, Michael D. and {Gangler}, Emmanuel and {Gawiser}, Eric and {Geary}, John C. and {Gee}, Perry and {Geha}, Marla and {Gessner}, Charles J.~B. and {Gibson}, Robert R. and {Gilmore}, D. Kirk and {Glanzman}, Thomas and {Glick}, William and {Goldina}, Tatiana and {Goldstein}, Daniel A. and {Goodenow}, Iain and {Graham}, Melissa L. and {Gressler}, William J. and {Gris}, Philippe and {Guy}, Leanne P. and {Guyonnet}, Augustin and {Haller}, Gunther and {Harris}, Ron and {Hascall}, Patrick A. and {Haupt}, Justine and {Hernandez}, Fabio and {Herrmann}, Sven and {Hileman}, Edward and {Hoblitt}, Joshua and {Hodgson}, John A. and {Hogan}, Craig and {Howard}, James D. and {Huang}, Dajun and {Huffer}, Michael E. and {Ingraham}, Patrick and {Innes}, Walter R. and {Jacoby}, Suzanne H. and {Jain}, Bhuvnesh and {Jammes}, Fabrice and {Jee}, M. James and {Jenness}, Tim and {Jernigan}, Garrett and {Jevremovi{\'c}}, Darko and {Johns}, Kenneth and {Johnson}, Anthony S. and {Johnson}, Margaret W.~G. and {Jones}, R. Lynne and {Juramy-Gilles}, Claire and {Juri{\'c}}, Mario and {Kalirai}, Jason S. and {Kallivayalil}, Nitya J. and {Kalmbach}, Bryce and {Kantor}, Jeffrey P. and {Karst}, Pierre and {Kasliwal}, Mansi M. and {Kelly}, Heather and {Kessler}, Richard and {Kinnison}, Veronica and {Kirkby}, David and {Knox}, Lloyd and {Kotov}, Ivan V. and {Krabbendam}, Victor L. and {Krughoff}, K. Simon and {Kub{\'a}nek}, Petr and {Kuczewski}, John and {Kulkarni}, Shri and {Ku}, John and {Kurita}, Nadine R. and {Lage}, Craig S. and {Lambert}, Ron and {Lange}, Travis and {Langton}, J. Brian and {Le Guillou}, Laurent and {Levine}, Deborah and {Liang}, Ming and {Lim}, Kian-Tat and {Lintott}, Chris J. and {Long}, Kevin E. and {Lopez}, Margaux and {Lotz}, Paul J. and {Lupton}, Robert H. and {Lust}, Nate B. and {MacArthur}, Lauren A. and {Mahabal}, Ashish and {Mandelbaum}, Rachel and {Markiewicz}, Thomas W. and {Marsh}, Darren S. and {Marshall}, Philip J. and {Marshall}, Stuart and {May}, Morgan and {McKercher}, Robert and {McQueen}, Michelle and {Meyers}, Joshua and {Migliore}, Myriam and {Miller}, Michelle and {Mills}, David J.},
        title = "{LSST: From Science Drivers to Reference Design and Anticipated Data Products}",
      journal = {\apj},
         year = 2019,
        month = mar,
       volume = {873},
       number = {2},
          eid = {111},
        pages = {111},
          doi = {10.3847/1538-4357/ab042c},
archivePrefix = {arXiv},
       eprint = {0805.2366},
 primaryClass = {astro-ph},
       adsurl = {https://ui.adsabs.harvard.edu/abs/2019ApJ...873..111I}
}

@INPROCEEDINGS{2013imputation,
	author={Agarwal, Shivam},
	booktitle={2013 International Conference on Machine Intelligence and Research Advancement}, 
	title={Data Mining: Data Mining Concepts and Techniques}, 
	year={2013},
	volume={},
	number={},
	pages={203-207},
	doi={10.1109/ICMIRA.2013.45}}

@article{luo_paper,
    author = {Luo, Zhijian and Tang, Zhirui and Chen, Zhu and Fu, Liping and Du, Wei and Zhang, Shaohua and Gong, Yan and Shu, Chenggang and Lu, Junhao and Li, Yicheng and Meng, Xian-Min and Zhou, Xingchen and Fan, Zuhui},
    title = {Imputation of missing photometric data and photometric redshift estimation for CSST},
    journal = {\mnras},
    volume = {531},
    number = {3},
    pages = {3539-3550},
    year = {2024},
    month = {06},
issn = {0035-8711},
    doi = {10.1093/mnras/stae1397},
    url = {https://doi.org/10.1093/mnras/stae1397},
    eprint = {https://academic.oup.com/mnras/article-pdf/531/3/3539/58198378/stae1397.pdf},
}

@INPROCEEDINGS{luken2021missing,
       author = {{Luken}, Kieran J. and {Padhy}, Rabina and {Wang}, X. Rosalind},
        title = "{Missing Data Imputation for Galaxy Redshift Estimation}",
    booktitle = {Machine Learning for Physical Sciences workshop at NeurIPS 2021},
         year = 2021,
        month = dec,
          eid = {1},
        pages = {1},
          doi = {10.48550/arXiv.2111.13806},
archivePrefix = {arXiv},
       eprint = {2111.13806},
 primaryClass = {astro-ph.IM},
       adsurl = {https://ui.adsabs.harvard.edu/abs/2021mlps.confE...1L}
}

@article{missing2022,
title = {Estimation of missing values in astronomical survey data: An improved local approach using cluster directed neighbor selection},
journal = {Information Processing and Management},
volume = {59},
number = {2},
pages = {102881},
year = {2022},
issn = {0306-4573},
doi = {https://doi.org/10.1016/j.ipm.2022.102881},
url = {https://www.sciencedirect.com/science/article/pii/S0306457322000127},
author = {Phimmarin Keerin and Tossapon Boongoen},
}

@article{Abdalla2011,
    author = {Abdalla, F. B. and Banerji, M. and Lahav, O. and Rashkov, V.},
    title = {A comparison of six photometric redshift methods applied to 1.5 million luminous red galaxies},
    journal = {\mnras},
    volume = {417},
    number = {3},
    pages = {1891-1903},
    year = {2011},
    month = {10},
    issn = {0035-8711},
    doi = {10.1111/j.1365-2966.2011.19375.x},
    url = {https://doi.org/10.1111/j.1365-2966.2011.19375.x},
    eprint = {https://academic.oup.com/mnras/article-pdf/417/3/1891/3805422/mnras0417-1891.pdf},
}

@article{tweak_templets,
doi = {10.1088/0004-637X/699/1/486},
url = {https://dx.doi.org/10.1088/0004-637X/699/1/486},
year = {2009},
month = {jun},
publisher = {The American Astronomical Society},
volume = {699},
number = {1},
pages = {486},
author = {Conroy, Charlie and Gunn, James E. and White, Martin},
title = {THE PROPAGATION OF UNCERTAINTIES IN STELLAR POPULATION SYNTHESIS MODELING. I. THE RELEVANCE OF UNCERTAIN ASPECTS OF STELLAR EVOLUTION AND THE INITIAL MASS FUNCTION TO THE DERIVED PHYSICAL PROPERTIES OF GALAXIES},
journal = {\apj},
}

@article{specz1,
    author = {Cole, Shaun and Percival, Will J. and Peacock, John A. and Norberg, Peder and Baugh, Carlton M. and Frenk, Carlos S. and Baldry, Ivan and Bland-Hawthorn, Joss and Bridges, Terry and Cannon, Russell and Colless, Matthew and Collins, Chris and Couch, Warrick and Cross, Nicholas J. G. and Dalton, Gavin and Eke, Vincent R. and de Propris, Roberto and Driver, Simon P. and Efstathiou, George and Ellis, Richard S. and Glazebrook, Karl and Jackson, Carole and Jenkins, Adrian and Lahav, Ofer and Lewis, Ian and Lumsden, Stuart and Maddox, Steve and Madgwick, Darren and Peterson, Bruce A. and Sutherland, Will and Taylor, Keith and The 2dFGRS Team},
    title = {The 2dF Galaxy Redshift Survey: power-spectrum analysis of the final data set and cosmological implications},
    journal = {\mnras},
    volume = {362},
    number = {2},
    pages = {505-534},
    year = {2005},
    month = {09},
    issn = {0035-8711},
    doi = {10.1111/j.1365-2966.2005.09318.x},
    url = {https://doi.org/10.1111/j.1365-2966.2005.09318.x},
    eprint = {https://academic.oup.com/mnras/article-pdf/362/2/505/6155670/362-2-505.pdf},
}

@article{specz2,
doi = {10.1086/510615},
url = {https://dx.doi.org/10.1086/510615},
year = {2007},
month = {mar},
publisher = {},
volume = {657},
number = {2},
pages = {645},
author = {Percival, Will J. and Nichol, Robert C. and Eisenstein, Daniel J. and Frieman, Joshua A. and Fukugita, Masataka and Loveday, Jon and Pope, Adrian C. and Schneider, Donald P. and Szalay, Alex S. and Tegmark, Max and Vogeley, Michael S. and Weinberg, David H. and Zehavi, Idit and Bahcall, Neta A. and Brinkmann, Jon and Connolly, Andrew J. and Meiksin, Avery},
title = {The Shape of the Sloan Digital Sky Survey Data Release 5 Galaxy Power Spectrum},
journal = {\apj},
}

@article{specz3,
    author = {Carrasco Kind, Matias and Brunner, Robert J.},
    title = {TPZ: photometric redshift PDFs and ancillary information by using prediction trees and random forests},
    journal = {\mnras},
    volume = {432},
    number = {2},
    pages = {1483-1501},
    year = {2013},
    month = {05},
    issn = {0035-8711},
    doi = {10.1093/mnras/stt574},
    url = {https://doi.org/10.1093/mnras/stt574},
    eprint = {https://academic.oup.com/mnras/article-pdf/432/2/1483/18463634/stt574.pdf},
}

@article{metric_outlier,
	author = {{Fotopoulou, S.} and {Paltani, S.}},
	title = {CPz: Classification-aided photometric-redshift estimation⋆},
	DOI= "10.1051/0004-6361/201730763",
	url= "https://doi.org/10.1051/0004-6361/201730763",
	journal = {\aap},
	year = 2018,
	volume = 619,
	pages = "A14",
}

@article{photoz1,
  title={Optical multicolors - A poor person's z machine for galaxies},
  author={David C. Koo},
  journal={\aj},
  year={1985},
  volume={90},
  pages={418-440},
  url={https://api.semanticscholar.org/CorpusID:120763096}
}

@ARTICLE{photoz2,
       author = {{Loh}, E.~D. and {Spillar}, E.~J.},
        title = "{Photometric Redshifts of Galaxies}",
      journal = {\apj},
         year = 1986,
        month = apr,
       volume = {303},
        pages = {154},
          doi = {10.1086/164062},
       adsurl = {https://ui.adsabs.harvard.edu/abs/1986ApJ...303..154L}
}

@article{Desprez2020,
author = {Desprez, Guillaume and Paltani, S. and Coupon, Jean and Almosallam, Ibrahim and Álvarez Ayllón, Alejandro and Amaro, Valeria and Brescia, Massimo and Brodwin, M. and Cavuoti, Stefano and Vicente-Albendea, J. and Fotopoulou, Sotiria and Hatfield, Peter and Hartley, W. and Ilbert, Olivier and Jarvis, M. and Longo, Giuseppe and Rau, M. and Saha, Ritika and Speagle, J. and Zucca, E.},
year = {2020},
month = {12},
pages = {A31},
title = {Euclid preparation: X. The Euclid photometric-redshift challenge},
volume = {644},
journal = {A\&A},
doi = {10.1051/0004-6361/202039403}
}

@INPROCEEDINGS{Venkatraman2015,
         author = {{Venkatraman}, Ramakrishnan and {Khaitan}, Siddhartha Kumar},
        title = "{A survey of techniques for designing and managing microgrids}",
    booktitle = {2015 IEEE Power \& Energy Society General Meeting},
         year = 2015,
        month = jul,
          eid = {996},
        pages = {996},
          doi = {10.1109/PESGM.2015.7286590},
       adsurl = {https://ui.adsabs.harvard.edu/abs/2015pesg.conf..996V}
}

@BOOK{Mo,
       author = {{Mo}, Houjun and {van den Bosch}, Frank C. and {White}, Simon},
        title = "{Galaxy Formation and Evolution}",
         year = 2010,
          doi = {10.1017/CBO9780511807244},
       adsurl = {https://ui.adsabs.harvard.edu/abs/2010gfe..book.....M}
}

@ARTICLE{evolution,
       author = {{Tasca}, L.~A.~M. and {Kneib}, J. -P. and {Iovino}, A. and {Le F{\`e}vre}, O. and {Kova{\v{c}}}, K. and {Bolzonella}, M. and {Lilly}, S.~J. and {Abraham}, R.~G. and {Cassata}, P. and {Cucciati}, O. and {Guzzo}, L. and {Tresse}, L. and {Zamorani}, G. and {Capak}, P. and {Garilli}, B. and {Scodeggio}, M. and {Sheth}, K. and {Zucca}, E. and {Carollo}, C.~M. and {Contini}, T. and {Mainieri}, V. and {Renzini}, A. and {Bardelli}, S. and {Bongiorno}, A. and {Caputi}, K. and {Coppa}, G. and {de La Torre}, S. and {de Ravel}, L. and {Franzetti}, P. and {Kampczyk}, P. and {Knobel}, C. and {Koekemoer}, A.~M. and {Lamareille}, F. and {Le Borgne}, J. -F. and {Le Brun}, V. and {Maier}, C. and {Mignoli}, M. and {Pello}, R. and {Peng}, Y. and {Perez Montero}, E. and {Ricciardelli}, E. and {Silverman}, J.~D. and {Vergani}, D. and {Tanaka}, M. and {Abbas}, U. and {Bottini}, D. and {Cappi}, A. and {Cimatti}, A. and {Ilbert}, O. and {Leauthaud}, A. and {Maccagni}, D. and {Marinoni}, C. and {McCracken}, H.~J. and {Memeo}, P. and {Meneux}, B. and {Oesch}, P. and {Porciani}, C. and {Pozzetti}, L. and {Scaramella}, R. and {Scarlata}, C.},
        title = "{The zCOSMOS redshift survey: the role of environment and stellar mass in shaping the rise of the morphology-density relation from z \raisebox{-0.5ex}\textasciitilde 1}",
      journal = {\aap},
         year = 2009,
        month = aug,
       volume = {503},
       number = {2},
        pages = {379-398},
          doi = {10.1051/0004-6361/200912213},
archivePrefix = {arXiv},
       eprint = {0906.4556},
 primaryClass = {astro-ph.CO},
       adsurl = {https://ui.adsabs.harvard.edu/abs/2009A\&A...503..379T}
}

@ARTICLE{evolution1,
       author = {{Conselice}, Christopher J.},
        title = "{The Evolution of Galaxy Structure Over Cosmic Time}",
      journal = {\araa},
         year = 2014,
        month = aug,
       volume = {52},
        pages = {291-337},
          doi = {10.1146/annurev-astro-081913-040037},
archivePrefix = {arXiv},
       eprint = {1403.2783},
 primaryClass = {astro-ph.GA},
       adsurl = {https://ui.adsabs.harvard.edu/abs/2014ARA\&A..52..291C}
}

@ARTICLE{spec_SN,
       author = {{Salvato}, Mara and {Ilbert}, Olivier and {Hoyle}, Ben},
        title = "{The many flavours of photometric redshifts}",
      journal = {Nat. Astron.},
         year = 2019,
        month = jun,
       volume = {3},
        pages = {212-222},
          doi = {10.1038/s41550-018-0478-0},
archivePrefix = {arXiv},
       eprint = {1805.12574},
 primaryClass = {astro-ph.GA},
       adsurl = {https://ui.adsabs.harvard.edu/abs/2019NatAs...3..212S}
}

@book{Missing_Data,
  title={Statistical Analysis with Missing Data, Third Edition},
  author={ Little, Roderick  and  Rubin, Donald },
  publisher={Statistical Analysis with Missing Data, Third Edition},
  year={2019},
}

@article{Feldmann,
    author = {Feldmann, R. and Carollo, C. M. and Porciani, C. and Lilly, S. J. and Capak, P. and Taniguchi, Y. and Fèvre, O. Le and Renzini, A. and Scoville, N. and Ajiki, M. and Aussel, H. and Contini, T. and McCracken, H. and Mobasher, B. and Murayama, T. and Sanders, D. and Sasaki, S. and Scarlata, C. and Scodeggio, M. and Shioya, Y. and Silverman, J. and Takahashi, M. and Thompson, D. and Zamorani, G.},
    title = {The Zurich Extragalactic Bayesian Redshift Analyzer and its first application: COSMOS},
    journal = {\mnras},
    volume = {372},
    number = {2},
    pages = {565-577},
    year = {2006},
    month = {09},
    issn = {0035-8711},
    doi = {10.1111/j.1365-2966.2006.10930.x},
    url = {https://doi.org/10.1111/j.1365-2966.2006.10930.x},
    eprint = {https://academic.oup.com/mnras/article-pdf/372/2/565/2937360/mnras0372-0565.pdf},
}

@article{csst,
doi = {10.3847/1538-4357/ab391e},
url = {https://dx.doi.org/10.3847/1538-4357/ab391e},
year = {2019},
month = {oct},
publisher = {The American Astronomical Society},
volume = {883},
number = {2},
pages = {203},
author = {Gong, Yan and Liu, Xiangkun and Cao, Ye and Chen, Xuelei and Fan, Zuhui and Li, Ran and Li, Xiao-Dong and Li, Zhigang and Zhang, Xin and Zhan, Hu},
title = {Cosmology from the Chinese Space Station Optical Survey (CSS-OS)},
journal = {\apj},
}

@article{csst1,
author = {Zhan, Hu},
year = {2021},
month = {04},
pages = {1290-1298},
title = {The wide-field multiband imaging and slitless spectroscopy survey to be carried out by the Survey Space Telescope of China Manned Space Program},
volume = {66},
journal = {Chinese Science Bulletin},
doi = {10.1360/TB-2021-0016}
}

@article{csst2,
doi = {10.1088/1674-4527/ac424e},
url = {https://dx.doi.org/10.1088/1674-4527/ac424e},
year = {2022},
month = {feb},
publisher = {National Astromonical Observatories, CAS and IOP Publishing},
volume = {22},
number = {2},
pages = {025019},
author = {Cao, Ye and Gong, Yan and Zheng, Zhen-Ya and Xu, Chun},
title = {Calibrating Photometric Redshift Measurements with the Multi-channel Imager (MCI) of the China Space Station Telescope (CSST)},
journal = {Res. Astron. Astrophys.},
}

@ARTICLE{1CSST,
       author = {{Zhan}, Hu},
        title = "{Consideration for a large-scale multi-color imaging and slitless spectroscopy survey on the Chinese space station and its application in dark energy research}",
      journal = {Scientia Sinica Physica, Mechanica \& Astronomica},
         year = 2011,
        month = jan,
       volume = {41},
       number = {12},
        pages = {1441},
          doi = {10.1360/132011-961},
       adsurl = {https://ui.adsabs.harvard.edu/abs/2011SSPMA..41.1441Z}
}

@ARTICLE{2CSST,
       author = {{Cao}, Ye and {Gong}, Yan and {Meng}, Xian-Min and {Xu}, Cong K. and {Chen}, Xuelei and {Guo}, Qi and {Li}, Ran and {Liu}, Dezi and {Xue}, Yongquan and {Cao}, Li and {Fu}, Xiyang and {Zhang}, Xin and {Wang}, Shen and {Zhan}, Hu},
        title = "{Testing photometric redshift measurements with filter definition of the Chinese Space Station Optical Survey (CSS-OS)}",
      journal = {\mnras},
         year = 2018,
        month = oct,
       volume = {480},
       number = {2},
        pages = {2178-2190},
          doi = {10.1093/mnras/sty1980},
archivePrefix = {arXiv},
       eprint = {1706.09586},
 primaryClass = {astro-ph.IM},
       adsurl = {https://ui.adsabs.harvard.edu/abs/2018MNRAS.480.2178C}
}

@ARTICLE{2018_Veronika,
       author = {{Veronika Dorogush}, Anna and {Ershov}, Vasily and {Gulin}, Andrey},
        title = "{CatBoost: gradient boosting with categorical features support}",
      journal = {arXiv e-prints},
         year = 2018,
        month = oct,
          doi = {10.48550/arXiv.1810.11363},
archivePrefix = {arXiv},
       eprint = {1810.11363},
 primaryClass = {cs.LG},
       adsurl = {https://ui.adsabs.harvard.edu/abs/2018arXiv181011363V}
}

@article{VanBuuren_2018,
 title={Flexible Imputation of Missing Data},
 volume={85},
 url={https://www.jstatsoft.org/index.php/jss/article/view/v085b04},
 doi={10.18637/jss.v085.b04},
 number={4},
 journal={Journal of Statistical Software, Book Reviews},
 author={Demirtas, Hakan},
 year={2018},
 pages={1–5}
}

@article{review,
   author = "Graham, John W.",
   title = "Missing Data Analysis: Making It Work in the Real World", 
   journal= "Annual Review of Psychology",
   year = "2009",
   volume = "60",
   number = "Volume 60, 2009",
   pages = "549-576",
   doi = "https://doi.org/10.1146/annurev.psych.58.110405.085530",
   url = "https://www.annualreviews.org/content/journals/10.1146/annurev.psych.58.110405.085530",
   publisher = "Annual Reviews",
   issn = "1545-2085",
   type = "Journal Article",
  }

@ARTICLE{2019DECaLS,
       author = {{Dey}, Arjun and {Schlegel}, David J. and {Lang}, Dustin and {Blum}, Robert and {Burleigh}, Kaylan and {Fan}, Xiaohui and {Findlay}, Joseph R. and {Finkbeiner}, Doug and {Herrera}, David and {Juneau}, St{\'e}phanie and {Landriau}, Martin and {Levi}, Michael and {McGreer}, Ian and {Meisner}, Aaron and {Myers}, Adam D. and {Moustakas}, John and {Nugent}, Peter and {Patej}, Anna and {Schlafly}, Edward F. and {Walker}, Alistair R. and {Valdes}, Francisco and {Weaver}, Benjamin A. and {Y{\`e}che}, Christophe and {Zou}, Hu and {Zhou}, Xu and {Abareshi}, Behzad and {Abbott}, T.~M.~C. and {Abolfathi}, Bela and {Aguilera}, C. and {Alam}, Shadab and {Allen}, Lori and {Alvarez}, A. and {Annis}, James and {Ansarinejad}, Behzad and {Aubert}, Marie and {Beechert}, Jacqueline and {Bell}, Eric F. and {BenZvi}, Segev Y. and {Beutler}, Florian and {Bielby}, Richard M. and {Bolton}, Adam S. and {Brice{\~n}o}, C{\'e}sar and {Buckley-Geer}, Elizabeth J. and {Butler}, Karen and {Calamida}, Annalisa and {Carlberg}, Raymond G. and {Carter}, Paul and {Casas}, Ricard and {Castander}, Francisco J. and {Choi}, Yumi and {Comparat}, Johan and {Cukanovaite}, Elena and {Delubac}, Timoth{\'e}e and {DeVries}, Kaitlin and {Dey}, Sharmila and {Dhungana}, Govinda and {Dickinson}, Mark and {Ding}, Zhejie and {Donaldson}, John B. and {Duan}, Yutong and {Duckworth}, Christopher J. and {Eftekharzadeh}, Sarah and {Eisenstein}, Daniel J. and {Etourneau}, Thomas and {Fagrelius}, Parker A. and {Farihi}, Jay and {Fitzpatrick}, Mike and {Font-Ribera}, Andreu and {Fulmer}, Leah and {G{\"a}nsicke}, Boris T. and {Gaztanaga}, Enrique and {George}, Koshy and {Gerdes}, David W. and {Gontcho}, Satya Gontcho A. and {Gorgoni}, Claudio and {Green}, Gregory and {Guy}, Julien and {Harmer}, Diane and {Hernandez}, M. and {Honscheid}, Klaus and {Huang}, Lijuan Wendy and {James}, David J. and {Jannuzi}, Buell T. and {Jiang}, Linhua and {Joyce}, Richard and {Karcher}, Armin and {Karkar}, Sonia and {Kehoe}, Robert and {Kneib}, Jean-Paul and {Kueter-Young}, Andrea and {Lan}, Ting-Wen and {Lauer}, Tod R. and {Le Guillou}, Laurent and {Le Van Suu}, Auguste and {Lee}, Jae Hyeon and {Lesser}, Michael and {Perreault Levasseur}, Laurence and {Li}, Ting S. and {Mann}, Justin L. and {Marshall}, Robert and {Mart{\'\i}nez-V{\'a}zquez}, C.~E. and {Martini}, Paul and {du Mas des Bourboux}, H{\'e}lion and {McManus}, Sean and {Meier}, Tobias Gabriel and {M{\'e}nard}, Brice and {Metcalfe}, Nigel and {Mu{\~n}oz-Guti{\'e}rrez}, Andrea and {Najita}, Joan and {Napier}, Kevin and {Narayan}, Gautham and {Newman}, Jeffrey A. and {Nie}, Jundan and {Nord}, Brian and {Norman}, Dara J. and {Olsen}, Knut A.~G. and {Paat}, Anthony and {Palanque-Delabrouille}, Nathalie and {Peng}, Xiyan and {Poppett}, Claire L. and {Poremba}, Megan R. and {Prakash}, Abhishek and {Rabinowitz}, David and {Raichoor}, Anand and {Rezaie}, Mehdi and {Robertson}, A.~N. and {Roe}, Natalie A. and {Ross}, Ashley J. and {Ross}, Nicholas P. and {Rudnick}, Gregory and {Safonova}, Sasha and {Saha}, Abhijit and {S{\'a}nchez}, F. Javier and {Savary}, Elodie and {Schweiker}, Heidi and {Scott}, Adam and {Seo}, Hee-Jong and {Shan}, Huanyuan and {Silva}, David R. and {Slepian}, Zachary and {Soto}, Christian and {Sprayberry}, David and {Staten}, Ryan and {Stillman}, Coley M. and {Stupak}, Robert J. and {Summers}, David L. and {Sien Tie}, Suk and {Tirado}, H. and {Vargas-Maga{\~n}a}, Mariana and {Vivas}, A. Katherina and {Wechsler}, Risa H. and {Williams}, Doug and {Yang}, Jinyi and {Yang}, Qian and {Yapici}, Tolga and {Zaritsky}, Dennis and {Zenteno}, A. and {Zhang}, Kai and {Zhang}, Tianmeng and {Zhou}, Rongpu and {Zhou}, Zhimin},
        title = "{Overview of the DESI Legacy Imaging Surveys}",
      journal = {\aj},
         year = 2019,
        month = may,
       volume = {157},
       number = {5},
          eid = {168},
        pages = {168},
          doi = {10.3847/1538-3881/ab089d},
archivePrefix = {arXiv},
       eprint = {1804.08657},
 primaryClass = {astro-ph.IM},
       adsurl = {https://ui.adsabs.harvard.edu/abs/2019AJ....157..168D}
}

@ARTICLE{2000HYPERZ,
	author = {{Bolzonella}, M. and {Miralles}, J. -M. and {Pell{\'o}}, R.},
	title = "{Photometric redshifts based on standard SED fitting procedures}",
	journal = {\aap},
	year = 2000,
	month = nov,
	volume = {363},
	pages = {476-492},
	doi = {10.48550/arXiv.astro-ph/0003380},
	archivePrefix = {arXiv},
	eprint = {astro-ph/0003380},
	primaryClass = {astro-ph},
	adsurl = {https://ui.adsabs.harvard.edu/abs/2000A&A...363..476B}
}

@ARTICLE{2006LePHARE,
	author = {{Ilbert}, O. and {Arnouts}, S. and {McCracken}, H.~J. and {Bolzonella}, M. and {Bertin}, E. and {Le F{\`e}vre}, O. and {Mellier}, Y. and {Zamorani}, G. and {Pell{\`o}}, R. and {Iovino}, A. and {Tresse}, L. and {Le Brun}, V. and {Bottini}, D. and {Garilli}, B. and {Maccagni}, D. and {Picat}, J.~P. and {Scaramella}, R. and {Scodeggio}, M. and {Vettolani}, G. and {Zanichelli}, A. and {Adami}, C. and {Bardelli}, S. and {Cappi}, A. and {Charlot}, S. and {Ciliegi}, P. and {Contini}, T. and {Cucciati}, O. and {Foucaud}, S. and {Franzetti}, P. and {Gavignaud}, I. and {Guzzo}, L. and {Marano}, B. and {Marinoni}, C. and {Mazure}, A. and {Meneux}, B. and {Merighi}, R. and {Paltani}, S. and {Pollo}, A. and {Pozzetti}, L. and {Radovich}, M. and {Zucca}, E. and {Bondi}, M. and {Bongiorno}, A. and {Busarello}, G. and {de La Torre}, S. and {Gregorini}, L. and {Lamareille}, F. and {Mathez}, G. and {Merluzzi}, P. and {Ripepi}, V. and {Rizzo}, D. and {Vergani}, D.},
	title = "{Accurate photometric redshifts for the CFHT legacy survey calibrated using the VIMOS VLT deep survey}",
	journal = {\aap},
	year = 2006,
	month = oct,
	volume = {457},
	number = {3},
	pages = {841-856},
	doi = {10.1051/0004-6361:20065138},
	archivePrefix = {arXiv},
	eprint = {astro-ph/0603217},
	primaryClass = {astro-ph},
	adsurl = {https://ui.adsabs.harvard.edu/abs/2006A&A...457..841I}
}

@ARTICLE{jiutian_2025,
       author = {{Han}, Jiaxin and {Li}, Ming and {Jiang}, Wenkang and {Chen}, Zhao and {Wang}, Huiyuan and {Wei}, Chengliang and {He}, Feihong and {He}, Jianhua and {Zhang}, Jiajun and {Liu}, Yu and {Cui}, Weiguang and {Gu}, Yizhou and {Guo}, Qi and {Jing}, Yipeng and {Kang}, Xi and {Li}, Guoliang and {Luo}, Xiong and {Luo}, Yu and {Pei}, Wenxiang and {Qiu}, Yisheng and {Tan}, Zhenlin and {Xie}, Lizhi and {Yang}, Xiaohu and {Yu}, Haoran and {Yu}, Yu and {Zhou}, Jiale},
        title = "{The Jiutian simulations for the CSST extra-galactic surveys}",
      journal = {Science China Physics, Mechanics, and Astronomy},
         year = 2025,
        month = aug,
       volume = {68},
       number = {10},
          eid = {109511},
        pages = {109511},
          doi = {10.1007/s11433-025-2712-1},
archivePrefix = {arXiv},
       eprint = {2503.21368},
 primaryClass = {astro-ph.CO},
       adsurl = {https://ui.adsabs.harvard.edu/abs/2025SCPMA..6809511H}
}

@article{SKlearn_2011,
  author  = {Fabian Pedregosa and Ga{{\"e}}l Varoquaux and Alexandre Gramfort and Vincent Michel and Bertrand Thirion and Olivier Grisel and Mathieu Blondel and Peter Prettenhofer and Ron Weiss and Vincent Dubourg and Jake Vanderplas and Alexandre Passos and David Cournapeau and Matthieu Brucher and Matthieu Perrot and {{\'E}}douard Duchesnay},
  title   = {Scikit-learn: Machine Learning in Python},
  journal = {Journal of Machine Learning Research},
  year    = {2011},
  volume  = {12},
  number  = {85},
  pages   = {2825-2830},
  url     = {http://jmlr.org/papers/v12/pedregosa11a.html}
}

@ARTICLE{wang_LLSimpute,
  author={Wang, Aiguo and Chen, Ye and An, Ning and Yang, Jing and Li, Lian and Jiang, Lili},
  journal={IEEE/ACM Transactions on Computational Biology and Bioinformatics}, 
  title={Microarray Missing Value Imputation: A Regularized Local Learning Method}, 
  year={2019},
  volume={16},
  number={3},
  pages={980-993},
  doi={10.1109/TCBB.2018.2810205}}

@article{ma2020_KNNimpute,
  author       = {Zongfang Ma and
                  Hongpeng Tian and
                  Zechao Liu and
                  Zuo{-}wei Zhang},
  title        = {A new incomplete pattern belief classification method with multiple
                  estimations based on {KNN}},
  journal      = {Appl. Soft Comput.},
  volume       = {90},
  pages        = {106175},
  year         = {2020},
  url          = {https://doi.org/10.1016/j.asoc.2020.106175},
  doi          = {10.1016/J.ASOC.2020.106175},
  bibsource    = {dblp computer science bibliography, https://dblp.org}
}

@book{vanbuuren2000mice,
  title={Multivariate imputation by chained equations: MICE V1. 0 user's manual},
  author={Van Buuren, Stef},
  year={2000},
  publisher={Leiden: TNO}
}

@ARTICLE{2017Schindler,
       author = {{Schindler}, Jan-Torge and {Fan}, Xiaohui and {McGreer}, Ian D. and {Yang}, Qian and {Wu}, Jin and {Jiang}, Linhua and {Green}, Richard},
        title = "{The Extremely Luminous Quasar Survey in the SDSS Footprint. I. Infrared-based Candidate Selection}",
      journal = {\apj},
         year = 2017,
        month = dec,
       volume = {851},
       number = {1},
          eid = {13},
        pages = {13},
          doi = {10.3847/1538-4357/aa9929},
archivePrefix = {arXiv},
       eprint = {1712.01205},
 primaryClass = {astro-ph.GA},
       adsurl = {https://ui.adsabs.harvard.edu/abs/2017ApJ...851...13S}
}

@inproceedings{Yoon2018Gain,
  title={Gain: Missing data imputation using generative adversarial nets},
  author={Yoon, Jinsung and Jordon, James and Schaar, Mihaela},
  booktitle={International conference on machine learning},
  pages={5689--5698},
  year={2018},
  organization={PMLR}
}

@ARTICLE{2024AJ_La,
       author = {{La Torre}, Valentina and {Sajina}, Anna and {Goulding}, Andy D. and {Marchesini}, Danilo and {Bezanson}, Rachel and {Pearl}, Alan N. and {Sodr{\'e}}, Laerte},
        title = "{Estimating Galaxy Parameters with Self-organizing Maps and the Effect of Missing Data}",
      journal = {\aj},
         year = 2024,
        month = jun,
       volume = {167},
       number = {6},
          eid = {261},
        pages = {261},
          doi = {10.3847/1538-3881/ad3821},
archivePrefix = {arXiv},
       eprint = {2403.18888},
 primaryClass = {astro-ph.GA},
       adsurl = {https://ui.adsabs.harvard.edu/abs/2024AJ....167..261L}
}

@ARTICLE{2023ApJ_Chartab,
       author = {{Chartab}, Nima and {Mobasher}, Bahram and {Cooray}, Asantha R. and {Hemmati}, Shoubaneh and {Sattari}, Zahra and {Ferguson}, Henry C. and {Sanders}, David B. and {Weaver}, John R. and {Stern}, Daniel K. and {McCracken}, Henry J. and {Masters}, Daniel C. and {Toft}, Sune and {Capak}, Peter L. and {Davidzon}, Iary and {Dickinson}, Mark E. and {Rhodes}, Jason and {Moneti}, Andrea and {Ilbert}, Olivier and {Zalesky}, Lukas and {McPartland}, Conor J.~R. and {Szapudi}, Istv{\'a}n and {Koekemoer}, Anton M. and {Teplitz}, Harry I. and {Giavalisco}, Mauro},
        title = "{A Machine-learning Approach to Predict Missing Flux Densities in Multiband Galaxy Surveys}",
      journal = {\apj},
         year = 2023,
        month = jan,
       volume = {942},
       number = {2},
          eid = {91},
        pages = {91},
          doi = {10.3847/1538-4357/acacf5},
archivePrefix = {arXiv},
       eprint = {2208.14781},
 primaryClass = {astro-ph.GA},
       adsurl = {https://ui.adsabs.harvard.edu/abs/2023ApJ...942...91C}
}

@ARTICLE{2025Euclid_PHZ,
  author = {{Euclid Collaboration} and {Tucci}, M. and {Paltani}, S. and {Hartley}, W.~G. and {Dubath}, F. and {Morisset}, N. and {Bolzonella}, M. and {Fotopoulou}, S. and {Tarsitano}, F. and {Saulder}, C. and {Pozzetti}, L. and {Enia}, A. and {Kang}, Y. and {Degaudenzi}, H. and {Saglia}, R. and {Salvato}, M. and {Ilbert}, O. and {Stanford}, S.~A. and {Roster}, W. and {Castander}, F.~J. and {Humphrey}, A. and {Landt}, H. and {Selwood}, M. and {Stevens}, G. and {Aghanim}, N. and {Altieri}, B. and {Amara}, A. and {Andreon}, S. and {Auricchio}, N. and {Aussel}, H. and {Baccigalupi}, C. and {Baldi}, M. and {Balestra}, A. and {Bardelli}, S. and {Battaglia}, P. and {Belikov}, A.~N. and {Bernardeau}, F. and {Biviano}, A. and {Bonchi}, A. and {Branchini}, E. and {Brescia}, M. and {Brinchmann}, J. and {Camera}, S. and {Ca{\~n}as-Herrera}, G. and {Capobianco}, V. and {Carbone}, C. and {Carretero}, J. and {Casas}, S. and {Castellano}, M. and {Castignani}, G. and {Cavuoti}, S. and {Chambers}, K.~C. and {Cimatti}, A. and {Colodro-Conde}, C. and {Congedo}, G. and {Conselice}, C.~J. and {Conversi}, L. and {Copin}, Y. and {Courbin}, F. and {Courtois}, H.~M. and {Cropper}, M. and {Da Silva}, A. and {De Lucia}, G. and {Di Giorgio}, A.~M. and {Dole}, H. and {Duncan}, C.~A.~J. and {Dupac}, X. and {Ealet}, A. and {Escoffier}, S. and {Fabricius}, M. and {Farina}, M. and {Farinelli}, R. and {Ferriol}, S. and {Finelli}, F. and {Fourmanoit}, N. and {Frailis}, M. and {Franceschi}, E. and {Fumana}, M. and {Galeotta}, S. and {George}, K. and {Gillard}, W. and {Gillis}, B. and {Giocoli}, C. and {G{\'o}mez-Alvarez}, P. and {Gracia-Carpio}, J. and {Granett}, B.~R. and {Grazian}, A. and {Grupp}, F. and {Haugan}, S.~V.~H. and {Hoar}, J. and {Hoekstra}, H. and {Holmes}, W. and {Hook}, I.~M. and {Hormuth}, F. and {Hornstrup}, A. and {Hudelot}, P. and {Jahnke}, K. and {Jhabvala}, M. and {Joachimi}, B. and {Keih{\"a}nen}, E. and {Kermiche}, S. and {Kiessling}, A. and {Kilbinger}, M. and {Kubik}, B. and {K{\"u}mmel}, M. and {Kunz}, M. and {Kurki-Suonio}, H. and {Le Boulc'h}, Q. and {Le Brun}, A.~M.~C. and {Le Mignant}, D. and {Liebing}, P. and {Ligori}, S. and {Lilje}, P.~B. and {Lindholm}, V. and {Lloro}, I. and {Mainetti}, G. and {Maino}, D. and {Maiorano}, E. and {Mansutti}, O. and {Marcin}, S. and {Marggraf}, O. and {Martinelli}, M. and {Martinet}, N. and {Marulli}, F. and {Massey}, R. and {Masters}, D.~C. and {Maurogordato}, S. and {McCracken}, H.~J. and {Medinaceli}, E. and {Mei}, S. and {Melchior}, M. and {Mellier}, Y. and {Meneghetti}, M. and {Merlin}, E. and {Meylan}, G. and {Mora}, A. and {Moresco}, M. and {Moscardini}, L. and {Nakajima}, R. and {Neissner}, C. and {Niemi}, S.-M. and {Nightingale}, J.~W. and {Padilla}, C. and {Pasian}, F. and {Pedersen}, K. and {Percival}, W.~J. and {Pettorino}, V. and {Pires}, S. and {Polenta}, G. and {Poncet}, M. and {Popa}, L.~A. and {Racca}, G.~D. and {Raison}, F. and {Rebolo}, R. and {Renzi}, A. and {Rhodes}, J. and {Riccio}, G. and {Romelli}, E. and {Roncarelli}, M. and {Rusholme}, B. and {Sakr}, Z. and {S{\'a}nchez}, A.~G. and {Sapone}, D. and {Sartoris}, B. and {Schewtschenko}, J.~A. and {Schirmer}, M. and {Schneider}, P. and {Secroun}, A. and {Sefusatti}, E. and {Seiffert}, M. and {Serrano}, S. and {Simon}, P. and {Sirignano}, C. and {Sirri}, G. and {Spurio Mancini}, A. and {Stanco}, L. and {Steinwagner}, J. and {Tallada-Cresp{\'\i}}, P. and {Taylor}, A.~N. and {Teplitz}, H.~I. and {Tereno}, I. and {Tessore}, N. and {Toft}, S. and {Toledo-Moreo}, R. and {Torradeflot}, F. and {Tutusaus}, I. and {Valentijn}, E.~A. and {Valenziano}, L. and {Valiviita}, J. and {Vassallo}, T. and {Verdoes Kleijn}, G. and {Veropalumbo}, A. and {Wang}, Y. and {Weller}, J. and {Zacchei}, A. and {Zamorani}, G. and {Zerbi}, F.~M. and {Zinchenko}, I.~A. and {Zucca}, E. and {Allevato}, V.},
  title   = {Euclid Quick Data Release (Q1). Photometric redshifts and physical properties of galaxies through the PHZ processing function},
  journal = {A\&A, accepted},
  year    = {2025},
  doi = {10.48550/arXiv.2503.15306},
archivePrefix = {arXiv},
       eprint = {2503.15306},
 primaryClass = {astro-ph.GA},
       adsurl = {https://ui.adsabs.harvard.edu/abs/2025arXiv250315306E}

}

@article{CSST_2025,
    author = {{CSST Collaboration} and {Gong}, Yan and {Miao}, Haitao and {Zhan}, Hu and {Li}, Zhao-Yu and {Shangguan}, Jinyi and {Li}, Haining and {Liu}, Chao and {Chen}, Xuefei and {Yuan}, Haibo and {Zhou}, Jilin and {Liu}, Hui-Gen and {Yu}, Cong and {Ji}, Jianghui and {Qi}, Zhaoxiang and {Liu}, Jiacheng and {Dai}, Zigao and {Wang}, Xiaofeng and {Zheng}, Zhenya and {Hao}, Lei and {Dou}, Jiangpei and {Ao}, Yiping and {Lin}, Zhenhui and {Zhang}, Kun and {Wang}, Wei and {Sun}, Guotong and {Li}, Ran and {Li}, Guoliang and {Xu}, Youhua and {Li}, Xinfeng and {Li}, Shengyang and {Wu}, Peng and {Zhang}, Jiuxing and {Wang}, Bo and {Bai}, Jinming and {Cai}, Yi-Fu and {Cai}, Zheng and {Cao}, Jie and {Chan}, Kwan Chuen and {Chang}, Jin and {Chen}, Xiaodian and {Chen}, Xuelei and {Chen}, Yuqin and {Chen}, Yun and {Cui}, Wei and {Dong}, Subo and {Du}, Pu and {Duan}, Wenying and {Fan}, Junhui and {Fan}, LuLu and {Fan}, Zhou and {Fan}, Zuhui and {Fang}, Taotao and {Fu}, Jianning and {Fu}, Liping and {Fu}, Zhensen and {Gao}, Jian and {Gu}, Shenghong and {Gu}, Yidong and {Guo}, Qi and {Han}, Zhanwen and {Hu}, Bin and {Huang}, Zhiqi and {Ho}, Luis C. and {Jiang}, Linhua and {Jiang}, Ning and {Jing}, Yipeng and {Kang}, Xi and {Kong}, Xu and {Li}, Cheng and {Li}, Chengyuan and {Li}, Di and {Li}, Jing and {Li}, Nan and {Li}, Yang A. and {Liao}, Shilong and {Lin}, Weipeng and {Liu}, Fengshan and {Liu}, Jifeng and {Liu}, Xiangkun and {Liu}, Zhuokai and {Mao}, Ruiqing and {Mao}, Shude and {Meng}, Xianmin and {Pang}, Xiaoying and {Peng}, Xiyan and {Peng}, Yingjie and {Shan}, Huanyuan and {Shen}, Juntai and {Shen}, Shiyin and {Shen}, Zhiqiang and {Shi}, Sheng-Cai and {Shi}, Yong and {Tan}, Siyuan and {Tian}, Hao and {Wang}, Jianmin and {Wang}, Jun-Xian and {Wang}, Xin and {Wang}, Yuting and {Wu}, Hong and {Wu}, Jingwen and {Wu}, Xuebing and {Xu}, Chun and {Xue}, Xiang-Xiang and {Xue}, Yongquan and {Yang}, Ji and {Yang}, Xiaohu and {Yao}, Qijun and {Yuan}, Fangting and {Yuan}, Zhen and {Zhang}, Jun and {Zhang}, Pengjie and {Zhang}, Tianmeng and {Zhang}, Wei and {Zhang}, Xin and {Zhao}, Gang and {Zhao}, Gongbo and {Zhong}, Hongen and {Zhong}, Jing and {Zhou}, Liyong and {Zhu}, Wei and {Zu}, Ying},
        title = "{Introduction to the Chinese Space Station Survey Telescope (CSST)}",
      journal = {Science China Physics, Mechanics, and Astronomy},
         year = 2026,
        month = jan,
       volume = {69},
       number = {3},
          eid = {239501},
        pages = {239501},
          doi = {10.1007/s11433-025-2809-0},
archivePrefix = {arXiv},
       eprint = {2507.04618},
 primaryClass = {astro-ph.IM},
       adsurl = {https://ui.adsabs.harvard.edu/abs/2026SCPMA..6939501C}
}

@article{Wei_2026_024001,
doi = {10.1088/1674-4527/ae20fe},
url = {https://doi.org/10.1088/1674-4527/ae20fe},
year = {2026},
month = {jan},
publisher = {National Astromonical Observatories, CAS and IOP Publishing},
volume = {26},
number = {2},
pages = {024001},
author = {Wei, Cheng-Liang and Li, Guo-Liang and Fang, Yue-Dong and Zhang, Xin and Luo, Yu and Tian, Hao and Liu, De-Zi and Meng, Xian-Ming and Ban, Zhang and Li, Xiao-Bo and Luo, Zun and Xian, Jing-Tian and Wang, Wei and Peng, Xi-Yan and Li, Nan and Li, Ran and Shao, Li and Zhang, Tian-Meng and Tang, Jing and Chen, Yang and Qi, Zhao-Xiang and Cao, Zi-Huang and Shan, Huan-Yuan and Nie, Lin and Yan, Zhaojun and Wang, Lei and He, Zizhao and Luo, Rui-Biao and Liu, Quan-Yu},
title = {Mock Observations for the CSST Mission: Main Surveys–An Overview of Framework and Simulation Suite},
journal = {Res. Astron. Astrophys.},
}

@article{Ban_2026,
doi = {10.1088/1674-4527/ae20fb},
url = {https://doi.org/10.1088/1674-4527/ae20fb},
year = {2026},
month = {jan},
publisher = {National Astromonical Observatories, CAS and IOP Publishing},
volume = {26},
number = {2},
pages = {024002},
author = {Ban, Zhang and Li, Xiao-Bo and Yang, Xun and Jiang, Yu-Xi and Ma, Hong-Cai and Wang, Wei and Lv, Jinguang and Wei, Cheng-Liang and Liu, De-Zi and Li, Guo-Liang and Liu, Chao and Li, Nan and Li, Ran and Wei, Peng},
title = {Mock Observations for the CSST Mission: End-to-end Performance Modeling of Optical System},
journal = {Res. Astron. Astrophys.},
}

@article{Wei_2026_024004,
doi = {10.1088/1674-4527/ae20ff},
url = {https://doi.org/10.1088/1674-4527/ae20ff},
year = {2026},
month = {jan},
publisher = {National Astromonical Observatories, CAS and IOP Publishing},
volume = {26},
number = {2},
pages = {024004},
author = {Wei, Cheng-Liang and Luo, Yu and Tian, Hao and Li, Ming and Qiu, Yi-Sheng and Li, Guo-Liang and Fang, Yue-Dong and Zhang, Xin and Liu, De-Zi and Li, Nan and Li, Ran and Shan, Huan-Yuan and Nie, Lin and He, Zizhao and Wang, Lei and Kang, Xi and Fan, Dongwei and Chen, Yang and Fu, Xiaoting and Liu, Chao},
title = {Mock Observations for the CSST Mission: Main Surveys–The Mock Catalog},
journal = {Res. Astron. Astrophys.},
}

@article{Xian_2026,
doi = {10.1088/1674-4527/ae20fc},
url = {https://doi.org/10.1088/1674-4527/ae20fc},
year = {2026},
month = {jan},
publisher = {National Astromonical Observatories, CAS and IOP Publishing},
volume = {26},
number = {2},
pages = {024005},
author = {Xian, Jing-Tian and Lin, Lin and Fang, Yue-Dong and Zhang, Xin and Xu, You-Hua and Meng, Xian-Min and Tian, Hao and Zhang, Tian-Yi and Ban, Zhang and Li, Guo-Liang and Xu, Shu-Yan and Wang, Wei},
title = {Mock Observations for the CSST Mission: Main Surveys—the Stray Light},
journal = {Res. Astron. Astrophys.},
}

@ARTICLE{2023_liu,
       author = {{Liu}, D.~Z. and {Meng}, X.~M. and {Er}, X.~Z. and {Fan}, Z.~H. and {Kilbinger}, M. and {Li}, G.~L. and {Li}, R. and {Schrabback}, T. and {Scognamiglio}, D. and {Shan}, H.~Y. and {Tao}, C. and {Ting}, Y.~S. and {Zhang}, J. and {Cheng}, S.~H. and {Farrens}, S. and {Fu}, L.~P. and {Hildebrandt}, H. and {Kang}, X. and {Kneib}, J.~P. and {Liu}, X.~K. and {Mellier}, Y. and {Nakajima}, R. and {Schneider}, P. and {Starck}, J.~L. and {Wei}, C.~L. and {Wright}, A.~H. and {Zhan}, H.},
        title = "{Potential scientific synergies in weak lensing studies between the CSST and Euclid space probes}",
      journal = {\aap},
         year = 2023,
        month = jan,
       volume = {669},
          eid = {A128},
        pages = {A128},
          doi = {10.1051/0004-6361/202243978},
archivePrefix = {arXiv},
       eprint = {2210.16341},
 primaryClass = {astro-ph.CO},
       adsurl = {https://ui.adsabs.harvard.edu/abs/2023A&A...669A.128L}
}

@ARTICLE{Agarwal2025TheBG,
       author = {{Agarwal}, Naman and {Dalal}, Siddhartha R. and {Misra}, Vishal},
        title = "{The Bayesian Geometry of Transformer Attention}",
      journal = {arXiv e-prints},
         year = 2025,
        month = dec,
          doi = {10.48550/arXiv.2512.22471},
archivePrefix = {arXiv},
       eprint = {2512.22471},
 primaryClass = {stat.ML},
       adsurl = {https://ui.adsabs.harvard.edu/abs/2025arXiv251222471A}
}

\begin{appendix} 
\onecolumn
\section{Imputation Performance Validation of the Model in the Output Catalog}\label{appendix1}

\begin{figure*}[ht!]
    \centering
    \begin{subfigure}{1\linewidth}
        \centering
        \includegraphics[width=1\textwidth]{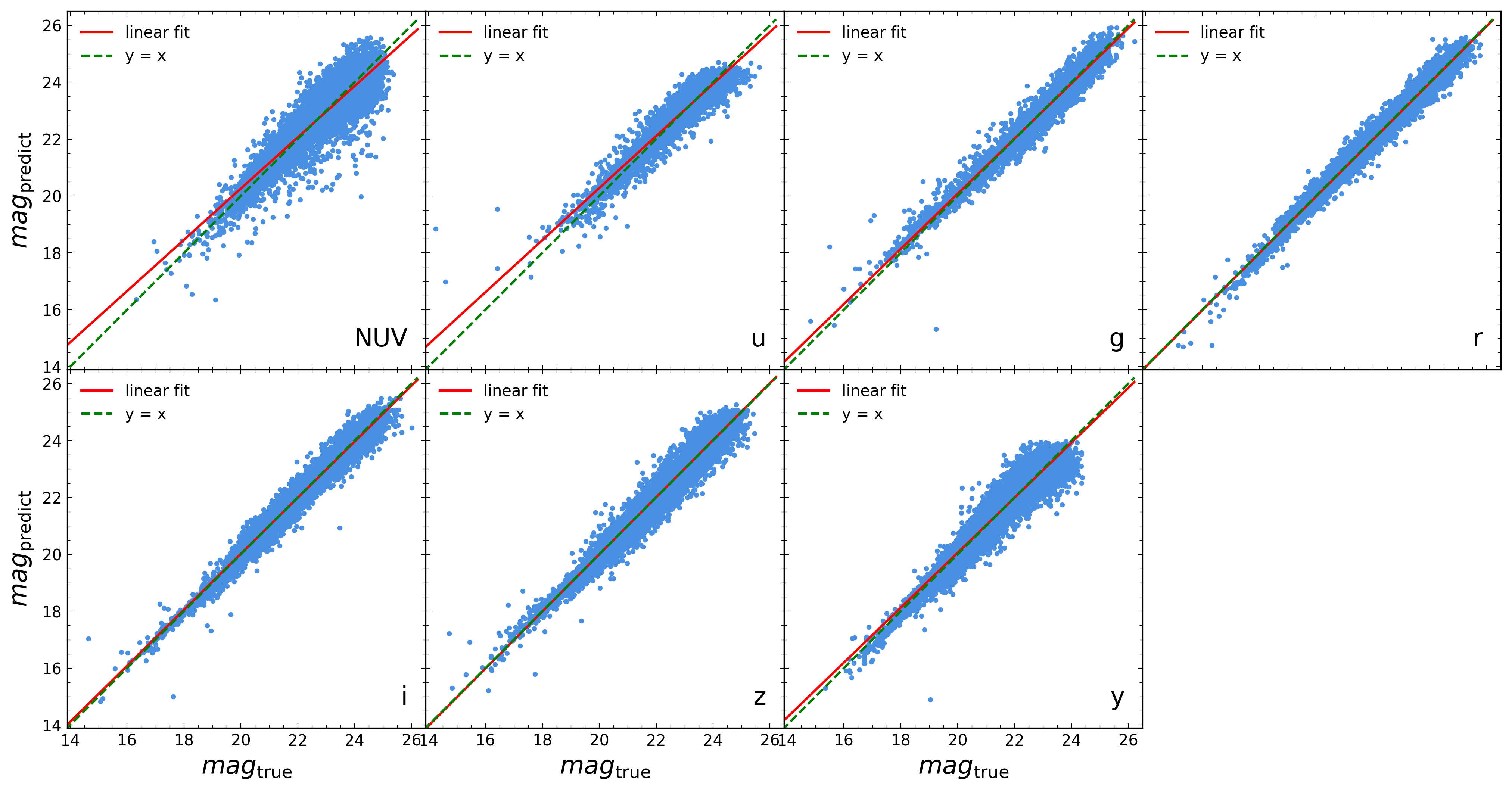}
	\end{subfigure}
    
    \begin{subfigure}{1\linewidth}
		\centering
        \includegraphics[width=1\textwidth]{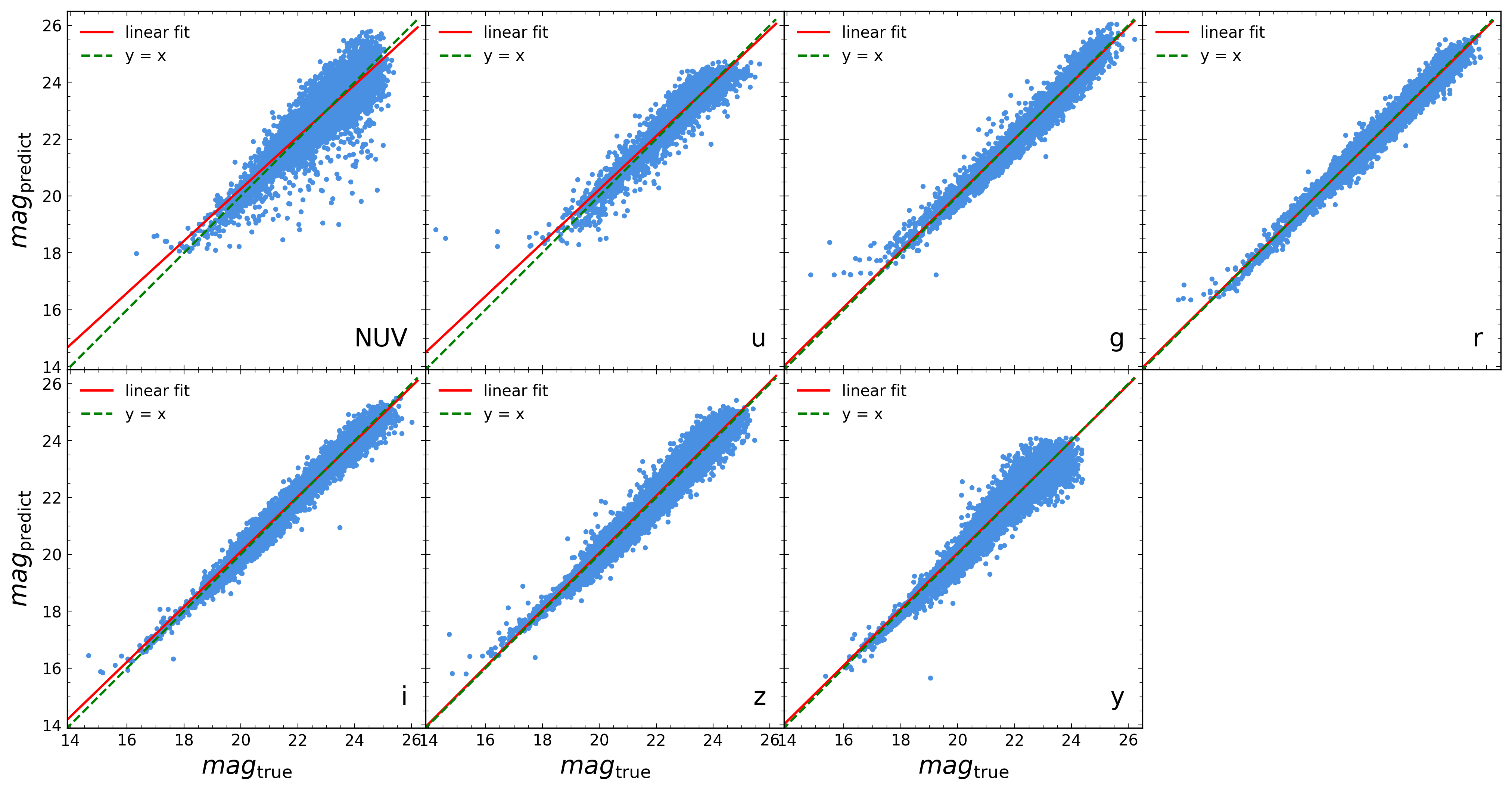}
	\end{subfigure}
    \caption{Same as Figure \ref{fig:input_imputation_dropband}, but for the output catalog. Panel a: Predicted versus true magnitudes (KNN). Panel b: Same as (a) but for the SAITS model. }\label{fig:output_imputation_dropband}
\end{figure*}

\end{appendix}

\end{document}